\documentclass[]{aa}  
\usepackage{txfonts}
\usepackage[colorlinks=true,citecolor=blue,urlcolor=blue,breaklinks=true]{hyperref}
\usepackage{graphicx}
\usepackage{placeins}           
\usepackage[normalem]{ulem}
\usepackage{rotating}    
\usepackage{booktabs}    
\usepackage{tabularx}    
\usepackage{array}       
\usepackage{threeparttable} 
\usepackage{url}         
\usepackage{caption}     

\begin{document}

\title{{Linking \it Fermi} blazars and radio galaxies through accretion and \\ jet radiation mechanisms}
\titlerunning{accretion and jet radiation mechanisms}
\authorrunning{Ye et al.}

   \author{Xu-Hong Ye\inst{1,2,3}
        \and Ranieri D. Baldi\inst{4}
        \and Wen-Xin Yang\inst{3,1,5}
        \and Jing-Tian Zhu\inst{6} 
        \and Denis Bastieri \inst{1,2,3}
        \and Rumen S. Bachev \inst{7}
        \and Anton A. Strigachev\inst{7}
        \and Jun-Hui Fan \inst{3,6}
        }

   \institute{Dipartimento di Fisica e Astronomia "G. Galilei", Università di Padova, Via F. Marzolo, 8, I-35131 Padova, Italy\\
             \email{denis.bastieri@unipd.it}
            \and Istituto Nazionale di Fisica Nucleare, Sezione di Padova, I-35131 Padova, Italy
            \and Centre for Astrophysics, Guangzhou University, Guangzhou 510006, People's Republic of China \\
            \email{fjh@gzhu.edu.cn}
            \and INAF - Istituto di Radioastronomia, via Gobetti 101, 40129 Bologna, Italy\\
            \email{ranieri.baldi@inaf.it}
            \and Astronomy Science and Technology Research Laboratory of Department of Education of Guangdong Province, Guangzhou 510006, People's Republic of China
            \and Wuhu Vocational Technical University, School of Network Engineering, Anhui, People's Republic of China 
            \and Institute of Astronomy and National Astronomical Observatory, Bulgarian Academy of Sciences, 72 Tsarigradsko shosse Blvd., 1784 Sofia, Bulgaria
            }

   \date{Received September 30, 20XX}

 
  \abstract
   {Based on the classical unification, blazars, namely BL Lacertae objects (BL Lacs) and flat-spectrum radio quasars (FSRQs), are believed to correspond with radio galaxies, when observed at small jet viewing angles.}
   {In this paper, we aim to compile a sample of \textit{Fermi} blazars [redshift, $z\in(0.002-4.313)$] and radio galaxies [$z\in(0.001-1.048)$] to provide new insights towards a unified accretion and ejection scenario between aligned and misaligned radio-loud active galactic nuclei (AGNs), by considering their optical emission-line classifications (low- and high-excitation radio galaxies, LERGs, HERGs), which are more representative of their accretion states.} 
   {We adopted the statistical analyses of accretion properties and high-energy beaming patterns for both \textit{Fermi} blazars and radio galaxies to investigate a unified accretion–ejection scenario.}
   {In the $\gamma$-ray luminosity–photon index plane, HERGs populate the region of higher luminosities and softer photon indices, akin to FSRQs, whereas LERGs fill at lower luminosities with harder photon indices, analogous to BL Lacs. This parallel segregation indicates that LERGs and HERGs represent the misaligned counterparts of BL Lacs and FSRQs, respectively. The unified picture is further supported by the Compton dominance–photon index diagram, where FSRQs and HERGs dominated by external Compton (EC) emissions are distinctly separated from BL Lacs and LERGs governed by synchrotron self-Compton (SSC) emissions. Similarly, the diagram of accretion rate versus $\gamma$-ray photon index reveals two distinct accretion–ejection states: a low-accretion-rate branch (BL Lacs and LERGs) is associated with the SSC model, and a high-accretion-rate branch (FSRQs and HERGs) is linked to the EC model. These results strongly strengthen the idea of a unified accretion and ejection paradigm between blazars and radio galaxies into two distinct states.}
  {}

   \keywords{Galaxies: active -- Accretion disks
                 --
                Galaxies: jets 
               }

\maketitle
   
\section{Introduction}\label{intro}
Blazars are a special subclass of radio-loud active galactic nuclei (RLAGNs) that exhibit observational properties dominated by relativistic jets, such as rapid variabilities, high luminosities and polarisations, apparent superluminal motions, radio core-dominance morphologies, or energetic high-energy emissions \citep{wills92,vermeulen94,fan96aas,homan.2021.apj,mooney.2021.apjs,fan.2021.apjs,abdollahi.apjs.2022.260,ajello.2022.apjs.263,liodakis2022nature,xiao2022mnras,yuan2023apjs,agudo25}. They come in two flavours based on the equivalent widths (EWs) of their emission lines: flat-spectrum radio quasars (FSRQs) have $\rm{EW} > 5\mathring{A}$, while BL Lacertae objects (BL Lacs) are defined by $\rm{EW} < 5\mathring{A}$ \citep{stickel.1991.apj.374}.

Radio galaxies, also a subclass of RLAGNs, are categorised into Fanaroff-Riley type I (FR~Is) and type II (FR~IIs) radio galaxies by the radio power and the extended morphology \citep{fanaroff.1974.mnras.167}. Recently, a very abundant class of nearby compact radio sources has been identified as FR type 0 (FR~0s) radio galaxies, which share similar core luminosities, but a factor of $\sim100-1000$ weaker jetted extended emission with respect to the FR~Is  \citep{ghisellini2011aipc,baldi23}. Radio galaxies (FR~0, FR~I, FR~II) are noted as misaligned AGNs because their relativistic jets are oriented away from the line of sight \citep{abdo.2010.apj.720.misagn}; when their jets align with observers, these radio galaxies are referred to as blazars (BL Lacs and FSRQs) in the classical unification paradigm \citep{ghisellini.1993.apj.407,urry.1995.pasp.107}. 

According to the emission-line properties, the population of radio galaxies can also be separated into the low-excitation radio galaxies (LERGs) and high-excitation radio galaxies (HERGs) \citep{best12mnras,heckman2014araa}, in which the former is associated with low-accretion rates,  typically, $\lambda_{\rm{Edd}}<0.01L_{\rm{Edd}}$\footnote{The $\lambda_{\rm Edd} = L_{\rm bol}/L_{\rm Edd}$ is a commonly-used proxy for the accretion rate and is a dimensionless Eddington ratio between the bolometric luminosity and the Eddington luminosity, $L_{\rm{Edd}}=1.38 \times 10^{38} (M/M_{\odot})$ erg/s, in which $M/M_{\odot}$ is the black hole mass in the unit of the solar mass (e.g., \citealt{raimundo09}).}, while the high-accretion rates feature the latter, approximately, $\lambda_{\rm{Edd}}\approx (0.01 - 0.1) L_{\rm{Edd}}$. The optical spectroscopic classes are more representative of the two accretion regimes onto the supermassive black holes (BHs) than the multiple radio morphological classes (e.g. \citealt{buttiglione10aa,mingo22,grandi25}). In general, FR 0s and FR Is are low-power LERGs with the central engine dominated by radiatively-inefficient, advective-dominated accretion flow (ADAF; \citealt{heckman2014araa,capetti.17.aa.fr1,baldi.18.aa.fr0,ye25aa}), but FR IIs are a heterogeneous population of high-power RLAGN with both LERGs and HERGs \citep{Tadhunter08,buttiglione10aa,baldi10b,capetti.17.aa.fr2,hardcastle20.newar,mingo22}, with the latter powered by a radiatively-efficient standard thin disc \citep{shakura73}.

The emission-line property not only serves as a classification criterion for both blazars and radio galaxies, but also as a reliable tracer for the accretion state, which is closely linked to the ejection mechanism \citep{ghisellini11mnras,sbarrato12mnras,ghi14nature,chen15aj}. \citet{sbarrato14mnras} found a tight correlation between accretion rate and jet power in blazars, highlighting a transition from radiatively efficient (e.g. FSRQs) to inefficient accretion (e.g. BL Lacs), each following distinct relations with their jet powers. Similar to the blazar case, the spectroscopy-based classification of the LERGs and HERGs also reflects a switch of the accretion regime and the jet power \citep{ghisellini01aa,hardcastle20.newar,boccardi2021aa,torresi22MnSAI,grandi25}.

The strong contributions of the relativistic jets to the spectral energy distributions (SEDs) shape a characteristic double-hump structure, either in blazars or in radio galaxies \citep{fossati98mnras,capetti.02.nar,ghisellini2010mnras,fan16,cerruti2020galaxies,yjh22apjs,ulgiati25}. In the leptonic model, the first hump of SEDs is attributed to the synchrotron radiation, from radio to soft X-ray bands \citep{finke2008apj}, while the second hump arises from the Inverse Compton (IC) radiation: either synchrotron self-Compton (SSC) or external Compton (EC) mechanisms due to different seed photon originations \citep{d&s1993ApJ,sik1994ApJ,bla2000ApJ,cerruti2020galaxies}. In the SSC model, synchrotron photons, produced within the jet as seed photons, are up-scattered by high-energy electrons, which is adopted to explain the $\gamma$-ray spectra of BL Lacs \citep{ghisellini2010mnras,zhao24apj,lian25}. In contrast, if the seed photons originate outside the jet, such as the broad-line region, dusty torus, accretion disc or cosmic microwave background (CMB) \citep{d&s1993ApJ, sik1994ApJ, bla2000ApJ, fan23apjs}, these external photons interacting with high-energy electrons in the jet, as EC process, are suited for explaining the $\gamma$-ray spectra in FSRQs
\citep{ghisellini2009mnras,chen2018apjs,cerruti2020galaxies}.  According to \citet{dermer1995apjl}, the beaming patterns are different between the SSC and EC models:  the SSC emissions go with the Doppler factors\footnote{The Doppler factor is defined by the Lorentz factor ($\Gamma$) and the jet viewing angle ($\theta$), $\delta=(\Gamma(1-\beta\cos\theta))^{-1}$, and $\Gamma=1/\sqrt{1-\beta^2}$, where $\beta$ is the bulk velocity in the unit of the speed of light ($\beta=v/c$).}($\delta$) in the observer frame as:  $f_{\rm{SSC}}\sim\delta_\gamma^{3+\alpha_{\gamma}}$, and the EC emissions go as  $f_{\rm{EC}}\sim\delta_\gamma^{4+2\alpha_{\gamma}}$, here $\alpha_{\gamma}$ is a $\gamma$-ray spectral index ($f_\nu \propto \nu^{-\alpha_{\gamma}}$). Therefore, the ratio of the beaming patterns between the EC and SSC models is $f_{\rm{EC}}/f_{\rm{SSC}}\propto \delta^{1+\alpha_{\gamma}}_{\gamma}$. The different high-energy beaming patterns between SSC and EC models help one to constrain the radiation models of the $\gamma$-ray sources \citep{dermer1995apjl,huang99apj,fan2013PASJ}.

Thanks to the \textit{Fermi} Large Area Telescope (\textit{Fermi}/LAT), one can observe a sample of thousands of blazars (e.g. \citealt{ajello.2022.apjs.263}) and study the high-energy leptonic or hadronic models. However, the hadronic model for blazars is argued to be ruled out with the evidence of the high optical to the X-ray polarisation ratio \citep{agudo25}. The hadronic processes for a single emission region and/or multiple emission regions expect that the polarisations in the X-ray band are comparable to those in the optical band \citep{zhc24apj}. Therefore, the low polarisation from the X-ray, compared to the optical band, favours a leptonic scenario where the low-energy photons are up-scattered to the X-ray band in BL Lacs. In addition, the multiwavelength polarisations for BL Lacs also support that the IC scattering from the electrons dominates at the X-ray energies, regardless of the jet compositions and emission models \citep{liodakis25aa}. These phenomena suggest that the leptonic jet model is a prominent scenario in blazars' $\gamma$-ray sky. 

From the 4 yrs of \textit{Fermi} observations, \citet{ghisellini2017mn} analysed a sample of 747 blazars, and showed that the observed $\gamma$-ray luminosities span a broad range from $10^{42}$ to $10^{50}$ erg s$^{-1}$. BL Lacs are typically found at lower redshifts ($z \lesssim 1$) and exhibit lower average luminosities with the harder $\gamma$-ray photon index ($\Gamma_{\gamma}=\alpha_{\gamma}+1$), while FSRQs are located at intermediate to high redshifts and display higher average luminosities with the softer $\gamma$-ray photon index. A \textit{Fermi} blazar sequence is proposed from a BL Lac population with the lower IC luminosity and the higher IC peak frequency to an FSRQ population with the higher IC luminosity and lower IC peak frequency. This phenomenon is explained by the different beaming patterns between the SSC and EC models \citep{fossati98mnras,ghisellini2017mn,boula26}. 

Besides the \textit{Fermi} blazars,  radio galaxies also appear in the $\gamma$-ray sky  \citep{abdo.2010.apj.720.misagn,ajello.2022.apjs.263}, and a comparable number between $\gamma$-ray FR Is and FR IIs is found in \citet{paliya24apj} and \citet{paliya25apj}. However, even though FR 0 is the dominant population of the local Universe, only a few have been detected or proposed as candidates by \textit{Fermi}/LAT \citep{grandi16mnras,baldi19,paliya21apjl,pannikkote23apj}.
Instead of the radio morphology and radio power, an accretion-ejection scenario seems to be more physically relevant to link radio galaxies and blazars: LERGs are suggested to be potential parent populations of the BL Lacs \citep{lai94,giommi2012mnras,chen15aj,mooney.2021.apjs}, in which FR 0 LERGs are highly connected to the nearby high-synchrotron peaked BL Lacs, sharing the similarities of the accretion disc, jet formation mechanisms, and clusters of the environments \citep{,massaro20apjl,ye25aa}. HERGs (mostly FR IIs) exhibit a luminous accretion disc and a powerful jet, similar to the FSRQs \citep{meyer2011apj,best12mnras,chen15aj,keenan21mn}.
Suppose that the distinct beaming patterns (SSC and EC models) help distinguish \textit{Fermi} BL Lacs from FSRQs. In that case, the same principle may also apply to separate \textit{Fermi} LERGs from HERGs, offering new insights into the \textit{Fermi} blazar–radio galaxy connection. This motivates the present paper, where we compiled a sample of \textit{Fermi} blazars and radio galaxies to discuss the high-energy leptonic models (SSC or EC), and proposed arguments in favour of a unified scenario between blazars and radio galaxies. Throughout the whole paper, a $\Lambda$-CMD cosmology, $\Omega_{\Lambda}\sim0.7$, $\Omega_{\rm{M}}\sim0.3$ and $H_0$ = 70 km s$^{-1}$ Mpc$^{-1}$, is considered.

\section{Samples}\label{sam}

\subsection{Blazars}
In this paper, we considered the BL Lacs and FSRQs using the fourth catalogue of AGNs detected by the \textit{Fermi}/LAT (4LAC) based on 12 years of data (Data Release 3; 4LAC-DR3), which includes 1667 blazars with available redshifts \citep{ajello.2022.apjs.263}.  Four of the 1667 blazars (4FGL J0601.3-7238, 4FGL J0654.0-4152, 4FGL J0719.7-4012, 4FGL J0828.3+4152) show extremely low $\gamma$-ray luminosities ($\log L_{\gamma}<40$ erg/s), and they could be misclassified as blazars. Therefore, these 4 sources are discarded. Meanwhile, 41 of 1663 blazars are cross-checked with the radio morphologies, radio core dominances and radio spectral index, which are re-classified as \textit{Fermi} radio galaxies in \citet{paliya24apj} and \citet{paliya25apj}. Therefore, the final blazar sample is 1622 sources with 838 BL Lacs and 784 FSRQs.

The redshift distribution of BL~Lacs spans $0.002 \leq z \leq 3.528$ with an average value of $\langle z_{\rm B} \rangle = 0.43 \pm 0.01$, whereas FSRQs range from $z=0.029$ to $z=4.313$ with a higher average redshift of $\langle z_{\rm F} \rangle = 1.20 \pm 0.02$. The $\gamma$-ray luminosities of 1622 blazars were computed using the integrated photon flux in the 1--100~GeV band, following the procedure described in \citet{yangwx2022apj} and \citet{xiao2022mnras}. The average $\gamma$-ray luminosity of FSRQs ($\log L_{\gamma}=46.09\pm0.03$ erg/s) is significantly higher than that of BL Lacs ($\log L_{\gamma}=44.92\pm0.03$ erg/s). The two populations exhibit statistically significant differences in both luminosity and photon-index distributions, as confirmed by Kolmogorov–Smirnov (K–S) tests with $p<10^{-4}$ for each parameter.

\begin{figure}[bpht]
    \centering
    \includegraphics[width=1.0\linewidth]{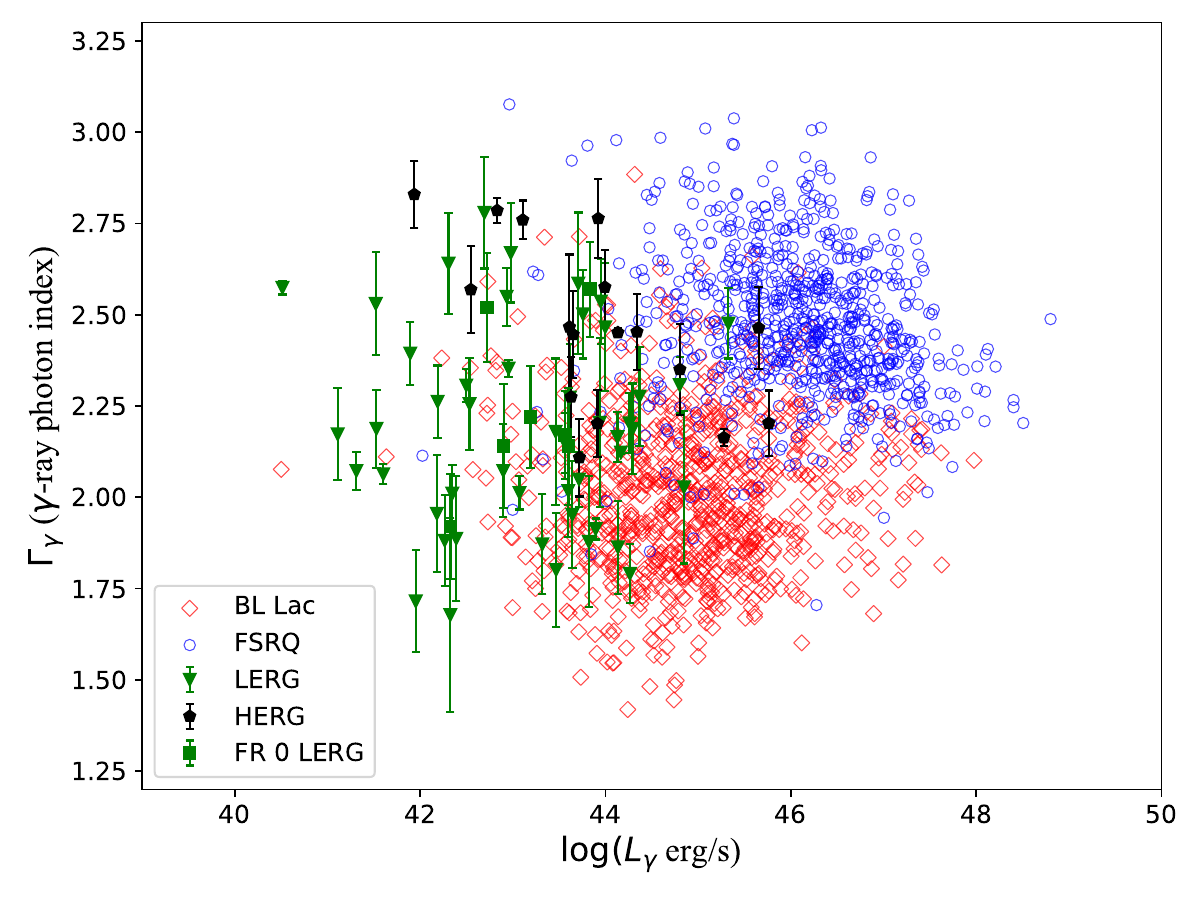}
    \caption{The plot of the relation between the $\gamma$-ray luminosities ($L_{\gamma}$) and $\gamma$-ray photon indices ($\Gamma_{\gamma}$). The green triangle is for LERGs, the green square is for FR 0 LERGs, and the black pentagram is for HERGs. BL Lacs are in red diamonds, while FSRQs are in blue circles. }
    \label{lum-index}
\end{figure}

\citet{ghisellini11mnras} investigated the relation between the $\gamma$-ray spectral index and $\gamma$-ray luminosity ($\Gamma_{\gamma}$--$L_{\gamma}$) and argued that the different loci of BL~Lacs and FSRQs in this plane reflect differences in jet power, ambient environment, and dominant high-energy emission mechanisms \citep{ghisellini2009mnras,ghisellini11mnras,sbarrato12mnras,sbarrato14mnras,ghisellini2017mn}. Our $\Gamma_{\gamma}$--$L_{\gamma}$ diagram (Fig. \ref{lum-index}) for the enlarged blazar sample exhibits a similar distribution to that reported by \citet{ghisellini11mnras}, further reinforcing the separation between FSRQs and BL~Lacs that arises from their different accretion regimes and high-energy radiation mechanisms \citep{sbarrato12mnras,paliya21apjs}.

\begin{table*}
\centering
\caption{Parameters for \textit{Fermi} blazars.}\label{tab:1}
\begin{tabular}{cccccccccc}
\hline
\hline 
 {4FGL Name} &  {$z$} &  {Class} &  $\Gamma_{\gamma}$ &  {$\log L_{\gamma}$} &   {$\log L_{\rm{bol}}$} & {CD} & $\log (L_{\rm{bol}}/L_{\rm{Edd}})$&Ref\\\
 (1)&(2)&(3)& (4) & (5) &(6) & (7) &(8) &(9)\\
\hline
J0001.5+2113	&	1.106	&	FSRQ	&	2.65	&	46.54	&	44.65	&	30.9	&	-1.03	&	P21	\\
J0003.2+2207	&	0.100	&	BL Lac	&	2.12	&	43.21	&	42.74	&	0.21	&	-3.50	&	P21	\\

 $\ldots$&$\ldots$&$\ldots$&$\ldots$&$\ldots$&$\ldots$&$\ldots$&$\ldots$ $\ldots$&$\ldots$\\
\hline
\end{tabular}
\medskip

\small
Notes: Col. (1) the source name; Col. (2) the redshift; Col. (3) the classification;  Col. (4) the $\gamma$-ray photon index; Col. (5) the $\gamma$-ray luminosity in the unit of erg/s; Col (6) the bolometric luminosity in the unit of erg/s;  Col. (7) the Compton dominance; Col. (8) the accretion rate; Col. (9) the reference for both the Compton dominance and the accretion rate, P21 for \citet{paliya21apjs}.

A portion of the Table is listed here for indication. The full Table is at the CDS.
\end{table*}

\begin{table*}[bpht]
\setlength{\tabcolsep}{3pt} 
\tiny
\centering
\caption{Parameters for \textit{Fermi} radio galaxies with the classification based on the BPT diagram.}\label{tab:2}
\begin{tabular}{ccccccccccccccc}
\hline
\hline 
 {4FGL Name} &Other name&  $z$ & $[\mathrm{N\,II}]/\mathrm{H}\alpha$ & $[\mathrm{O\,III}]/\mathrm{H}\beta$ & {Class} & Ref  & $\Gamma_{\gamma}$ &  {$\log L_{\gamma}$} &  {CD}  & Ref & {$\log L_{\rm{bol}}$} & Ref & $\log (L_{\rm{bol}}/L_{\rm{Edd}})$ &Ref\\\
 (1)&(2)&(3)& (4) & (5) &(6) & (7) &(8) & (9)&(10) &(11)&(12)&(13)& (14) & (15)\\
\hline
J0003.2+2207 	&	LEDA 1663156	&	0.099	&		&		&	LERG (FR0)	&	P23	&	2.14	&	43.60 	&		&		&	42.98	&	P23	&	-3.14	&	P23	\\
J0014.2+0854	&	MS 0011.7+0837	&	0.163	&	0.17	&	-0.30	&	LERG	&	SDSS	&	2.50	&	43.76 	&	0.43	&	P21	&	43.74	&	SDSS	&	-3.21	&	 S-T02	\\
J0038.7-0204	&	3C 17	&	0.22	&		&		&	HERG	&	B10	&	2.76	&	43.92 	&	5.89	&	C23	&	45.53	&	B10	&	-1.28	&	S07	\\
J0049.0+2252	&	PKS J0049+2253	&	0.264	&	0.38	&	0.35	&	LERG	&	SDSS	&	2.28	&	44.37 	&	0.62	&	P21	&	44.17	&	SDSS	&	-2.69	&	 S-T02	\\
J0154.3-0236	&	LEDA 144405	&	0.082	&		&		&	LERG (FR0)	&	P23	&	2.17	&	43.56 	&		&		&	43.94	&	P23	&	-3.58	&	P23	\\
J0308.4+0407	&	NGC 1218	&	0.029	&		&		&	LERG	&	B10	&	2.01	&	43.07 	&	0.45	&	C23	&	42.95	&	B10	&	-3.82	&	M14-T02	\\
J0312.4-3221	&	NVSS J031234-322315	&	0.067	&		&		&	LERG (FR0)	&	P23	&	2.22	&	43.19 	&		&		&	44.08	&	P23	&	-2.75	&	P23	\\
J0316.8+4120	&	IC 310	&	0.018	&	-0.03	&	-0.22	&	LERG	&	SDSS	&	1.88	&	42.26 	&	0.03	&	C23	&	42.83	&	M14	&	-3.61	&	S-T02	\\
J0319.8+4130	&	NGC 1275	&	0.018	&		&		&	LERG	&	B10	&	2.12	&	44.17 	&	3.02	&	C23	&	43.14	&	B10	&	-3.45	&	M14-T02	\\
J0418.2+3807	&	3C 111	&	0.048	&		&		&	HERG	&	B10	&	2.76	&	43.11 	&	0.41	&	C23	&	44.30	&	B10	&	-2.19	&	E03	\\
J0519.6-4544	&	Pictor A	&	0.035	&		&		&	HERG	&	F85	&	2.57	&	42.55 	&	1.51	&	C23	&	44.44	&	H16	&	-1.27	&	L06	\\
J0522.9-3628	&	PKS 0521-36	&	0.056	&		&		&	HERG	&	R87	&	2.45	&	44.13 	&	0.63	&	P21	&		&		&		&		\\
J0708.9+4839	&	NGC 2329	&	0.019	&		&		&	LERG	&	K08	&	1.72	&	41.95 	&	0.32	&	C23	&	42.66	&	E10	&	-3.85	&	M14-T02	\\
J0758.7+3746	&	NGC 2484	&	0.042	&	0.40	&	0.18	&	LERG	&	SDSS	&	2.26	&	42.53 	&	0.03	&	C23	&	42.26	&	SDSS	&	-4.49	&	S-T02	\\
J0829.0+1755	&	TXS 0826+180	&	0.089	&	0.17	&	-0.21	&	LERG	&	SDSS	&	2.15	&	43.56 	&	0.19	&	P21	&	42.88	&	SDSS	&	-3.42	&	 S-T02	\\
J1116.6+2915	&	B2 1113+29	&	0.046	&	0.53	&	0.47	&	LERG	&	SDSS	&	1.68	&	42.32 	&	0.30	&	C23	&	41.90	&	SDSS	&	-4.84	&	S-T02	\\
J0931.9+6737	&	NGC 2892	&	0.022	&		&		&	LERG	&	N03	&	2.31	&	42.50 	&	0.02	&	C23	&		&		&		&		\\
J1139.6+1149	&	4C +12.42	&	0.081	&	0.14	&	0.15	&	LERG	&	SDSS	&	2.78	&	42.69 	&		&		&	42.38	&	SDSS	&	-4.36	&	 S-T02	\\
J1144.9+1937	&	3C 264	&	0.021	&	0.16	&	-0.19	&	LERG	&	B10	&	2.01	&	42.35 	&	0.04	&	C23	&	42.74	&	B10	&	-3.99	&	M14-T02	\\
J1202.4+4442	&	B3 1159+450	&	0.297	&	-0.30	&	-0.65	&	LERG	&	SDSS	&	2.47	&	44.00 	&	0.69	&	P21	&	43.58	&	SDSS	&	-2.89	&	 S-T02	\\
J1212.1+6412	&	LEDA 2665658	&	0.108	&	0.06	&	0.06	&	LERG (FR0)	&	P23	&	2.57	&	43.83 	&		&		&	44.05	&	P23	&	-2.49	&	P23	\\
J1149.0+5924	&	NGC 3894	&	0.011	&		&		&	LERG	&	GS04	&	2.19	&	41.53 	&	0.04	&	C23	&	43.63	&	B21	&	-3.78	&	 B21	\\
J1216.1+0930	&	TXS 1213+097	&	0.093	&	0.14	&	-0.23	&	LERG	&	SDSS	&	2.05	&	43.71 	&	0.30	&	P21	&	43.23	&	SDSS	&	-3.61	&	 S-T02	\\
J1226.9+6405	&	GB6 J1226+6406	&	0.11	&	0.17	&	-0.17	&	LERG	&	SDSS	&	2.67	&	42.98 	&		&		&	42.28	&	SDSS	&	-4.42	&	 S-T02	\\
J1233.6+5027	&	TXS 1231+507	&	0.206	&	-0.14	&	0.25	&	LERG	&	SDSS	&	2.20	&	44.26 	&	0.59	&	P21	&	43.87	&	SDSS	&	-2.78	&	 S-T02	\\
J1306.3+1113	&	TXS 1303+114	&	0.085	&	0.38	&	0.14	&	LERG	&	C17a	&	1.87	&	43.32 	&	0.09	&	C23	&	43.48	&	C17a	&	-3.23	&	C17a	\\
J1230.8+1223	&	M 87	&	0.004	&		&		&	LERG	&	B10	&	2.06	&	41.60 	&		&		&	42.53	&	B10	&	-4.55	&	M14-T02	\\
J1326.2+4115	&	B3 1323+415	&	0.309	&	-0.04	&	0.47	&	HERG	&	SDSS	&	2.45	&	44.34 	&		&		&	43.68	&	SDSS	&	-2.28	&	 S-T02	\\
J1327.0+3154	&	B2 1325+32	&	0.239	&	-0.06	&	0.13	&	LERG	&	SDSS	&	2.20	&	43.94 	&		&		&		&		&		&		\\
J1325.5-4300	&	Cen A	&	0.001	&		&		&	LERG	&	SM98	&	2.57	&	40.51 	&		&		&	42.98	&	Bo21	&	-2.87	&	C09	\\
J1341.2+3958	&	SDSS J134105.10+395945.4	&	0.172	&	0.05	&	-0.38	&	LERG	&	SDSS	&	1.79	&	44.26 	&	0.15	&	P21	&	43.63	&	SDSS	&	-3.15	&	 S-T02	\\
J1342.7+0505	&	4C +05.57	&	0.136	&	0.07	&	0.63	&	HERG	&	SDSS	&	2.20	&	43.91 	&	0.21	&	P21	&	43.91	&	SDSS	&	-2.69	&	 S-T02	\\
J1330.1-3818	&	Tol 1326-379	&	0.028	&		&		&	LERG	&	G16	&	2.18	&	43.46 	&		&		&	44.10	&	G16	&	-2.31	&	G16	\\
J1340.1+3857	&	NVSS J133849+385111	&	0.246	&		&		&	LERG	&	B11	&	2.54	&	43.95 	&	0.50	&	P21	&		&		&		&		\\
J1352.6+3133	&	3C 293	&	0.045	&	0.02	&	-0.05	&	LERG	&	SDSS	&	2.64	&	42.31 	&		&		&	42.56	&	SDSS	&	-3.75	&	 S-T02	\\
J1402.6+1600	&	4C +16.39	&	0.244	&	-0.06	&	0.04	&	LERG	&	SDSS	&	2.19	&	44.29 	&	0.19	&	P21	&	44.41	&	SDSS	&	-1.89	&	 S-T02	\\
J1443.1+5201	&	3C 303	&	0.141	&	-0.05	&	0.76	&	HERG	&	B10	&	2.11	&	43.72 	&	0.47	&	C23	&	45.28	&	B10	&	-1.86	&	Hu16	\\
J1512.2+0202	&	PKS 1509+022	&	0.219	&	0.00	&	0.55	&	HERG	&	SDSS	&	2.16	&	45.28 	&	1.78	&	P21	&	44.44	&	SDSS	&	-1.97	&	 S-T02	\\
J1516.5+0015	&	PKS 1514+00	&	0.052	&	0.08	&	0.42	&	LERG	&	C17b	&	2.55	&	42.93 	&	0.47	&	C23	&	44.49	&	C17b	&	-2.42	&	C17b	\\
J1518.6+0614	&	TXS 1516+064	&	0.102	&	-0.07	&	-0.54	&	LERG	&	C17a	&	1.80	&	43.47 	&	0.74	&	C23	&	43.40	&	C17a	&	-3.51	&	C17a	\\
J1521.1+0421	&	PKS B1518+045	&	0.052	&	-0.13	&	-0.08	&	LERG	&	C17a	&	2.07	&	42.90 	&	0.56	&	C23	&	43.09	&	C17a	&	-3.92	&	M14	\\
J1530.3+2709	&	LEDA 55267	&	0.033	&	0.07	&	-0.86	&	LERG (FR0)	&	Ba18	&	1.92	&	42.33 	&		&		&	43.21	&	Ba18	&	-3.07	&	Ba18	\\
J1541.1+3451	&	FIRST J154058.6+345224	&	0.233	&	0.14	&	0.49	&	LERG	&	SDSS	&	2.59	&	43.71 	&		&		&	43.50	&	SDSS	&	-3.43	&	 S-T02	\\
J1556.1+2812	&	NVSS J155611+281134	&	0.208	&	0.56	&	-0.05	&	LERG	&	SDSS	&	1.88	&	43.82 	&		&		&	43.45	&	SDSS	&	-2.93	&	 S-T02	\\
J1606.4+1814	&	NGC 6061	&	0.036	&	0.46	&	-0.55	&	LERG	&	SDSS	&	1.89	&	42.39 	&		&		&	41.61	&	SDSS	&	-5.20	&	 S-T02	\\
J1628.8+2529	&	LEDA 58287	&	0.04	&		&		&	LERG	&	SDSS	&	1.96	&	42.18 	&		&		&	43.31	&	P23	&	-3.01	&	P23	\\
J1644.2+4546	&	B3 1642+458	&	0.225	&	0.00	&	-0.26	&	LERG	&	SDSS	&	1.86	&	44.13 	&	0.08	&	P21	&	43.80	&	SDSS	&	-3.03	&	 S-T02	\\
J2326.9-0201	&	PKS 2324-02	&	0.188	&	-0.05	&	0.62	&	HERG	&	SDSS	&	2.58	&	43.99 	&	3.24	&	C23	&	45.44	&	SDSS	&	-1.46	&	 S-T02	\\
J1612.2+2828	&	TXS 1610+285	&	0.053	&		&		&	LERG (FR0)	&	P23	&	2.14	&	42.90 	&	0.05	&	P21	&	43.44	&	P23	&	-2.97	&	P23	\\
J1612.4-0554	&	LEDA 1038366	&	0.029	&		&		&	LERG (FR0)&	P23	&	2.52	&	42.72 	&		&		&	42.52	&	P23	&	-4.23	&	P23	\\
J2330.4+1230	&	TXS 2327+121	&	0.144	&	0.17	&	0.45	&	LERG	&	SDSS	&	2.27	&	43.61 	&		&		&	43.85	&	SDSS	&	-2.89	&	 S-T02	\\
J1724.2-6501	&	NGC 6328	&	0.014	&		&		&	LERG	&	F85	&	2.53	&	41.52 	&	0.08	&	C23	&	42.00	&	E10	&	-4.73	&	W10	\\
J1824.7-3243	&	PKS 1821-327	&	0.355	&		&		&	HERG	&	M09	&	2.35	&	44.80 	&		&		&		&		&		&		\\
J2302.8-1841	&	PKS 2300-18	&	0.128	&		&		&	HERG	&	R87	&	2.27	&	43.63 	&	1.51	&	C23	&	45.58	&	K17	&	-0.97	&	W04	\\
									
\hline
\end{tabular}
\medskip

\small
Notes: Col. (1) the source name; Col. (2) the other name; Col. (3) redshift; Col. (4)-(5) logarithm of the emission-line ratios [N II 6584$\mathring{A}$/H${\alpha}$] and [O III 5007$\mathring{A}$/H${\beta}$]; Col. (6)-(7) classification and its reference; Col. (8) $\gamma$-ray photon index; Col. (9) $\gamma$-ray luminosities in the unit of erg/s; Col. (10)-(11) the Compton dominance and its reference; Col. (12)-(13) the bolometric luminosity and its reference; Col.(14)-(15) the accretion rate and its reference. If the stellar velocity dispersions from the SDSS  and/or HyperLeda website (\url{http://atlas.obs-hp.fr/hyperleda}, \citealt{makarov14AA}) are available, then the BH masses are computed from the relation between the BH masses and velocity dispersion \citep{tremaine2002ApJ} (S-T02 or M14-T02); otherwise, the BH masses are obtained from the literature. References: 
F85 for \citet{filip1985ApJ}, R87 for \citet{robinso87}, 
SM98 for \citet{SM98},  
T02 for \citet{tremaine2002ApJ}, E03 for \citet{eracleous03apj},  
N03 for \citet{noel2003ApJS}, W04 for \citet{wu&liu2004ApJ}, GS04 for \citet{g&s2004A&A}, L06 for \citet{lewis06apj}, S07 for \citet{sikora07},  
K08 for \citet{kolla2008A&A}, 
C09 for \citet{cappellari09}, 
M09 for \citet{m09},
B10 for \citet{buttiglione10aa}, 
E10 for \citet{evans10}, 
W10 for \citet{willett2010ApJ}, 
B11 for \citet{b11},
M14 for \citet{makarov14AA}, 
G16 for \citet{grandi16mnras}, H16 for \citet{hardcastle16}, Hu16 for \citet{hu2016RAA}, 
C17a for \citet{capetti.17.aa.fr1}, C17b for \citet{capetti.17.aa.fr2}, K17 for \citet{koss17},
Ba18 for \citet{baldi.18.aa.fr0},
B21 for \citet{balasu2021ApJ}, Bo21 for \citet{borkar2021MNRAS}, P21 for \citet{paliya21apjs}; C23 for \citet{chen23apj_sed},
P23 for \citet{pannikkote23apj}, SDSS for Sloan Digital Sky Survey (\url{www.sdss.org}, \citealt{sdss_cite}).
\end{table*}
\subsection{Radio Galaxies}
The classical classification of radio galaxies (FR I or FR II) is based on their jet extended morphology, which is found to be related to the radio power \citep{fanaroff.1974.mnras.167}. However, this  observational radio dichotomy may hide the physical nature of their central engines among the subclasses of radio galaxies \citep{hardcastle07}. LERGs and HERGs are separated by the emission-line properties as indicated by the accretion-ejection paradigm \citep{heckman2014araa}. The accretion-ejection diagram between LERGs and HERGs is a more physical classification criterion for radio galaxies, and the differences are explained by the changes in accretion rate from inefficient to efficient disc \citep{ghisellini01aa,hardcastle07,buttiglione10aa,best12mnras,heckman2014araa,mingo22}.

From the 14 years of observations of the fourth \textit{Fermi} catalogue (4FGL-DR4), 53 $\gamma$-ray radio galaxies are reported \citep{abdollahi.apjs.2022.260}. 
Recently, \citet{paliya24apj} and \citet{paliya25apj} systematically examined the radio morphologies of $\gamma$-ray AGNs and applied three diagnostic criteria to identify misaligned $\gamma$-ray radio galaxies: 
(i) optical spectra dominated by narrow emission lines and/or galaxy spectral features; 
(ii) low radio core dominance ($R<1$), and 
(iii) steep radio spectra ($\alpha_{\rm{r}}>0.5$). 
Using these diagnostics, they identified 113 additional \textit{Fermi} radio galaxies that satisfy, at least, two of the three criteria, bringing the total number of \textit{Fermi} radio galaxies to 166.

We first cross-matched these 166 sources with the literature and identified 23 radio galaxies with optical classifications as LERGs or HERGs\footnote{We group narrow-lined HERGs and broad-lined radio galaxies (BLRGs) together, as BLRGs represent type-I HERGs whose broad permitted lines are visible in optical spectra \citep[e.g.][]{morganti1999AA,buttiglione10aa,baldi2013aa,mingo14mnras,sadler2014MNRAS,ineson2015MNRAS,macconi2020mnras,boccardi2021aa}.} based on the BPT (Baldwin–Phillips–Terlevich) diagram \citep[e.g.][]{buttiglione10aa}. The classifications are listed in Col. (6) with the reference in Col. (7) in Table \ref{tab:2}.

\begin{figure}[bpht]
    \centering
    \includegraphics[width=1.0\linewidth]{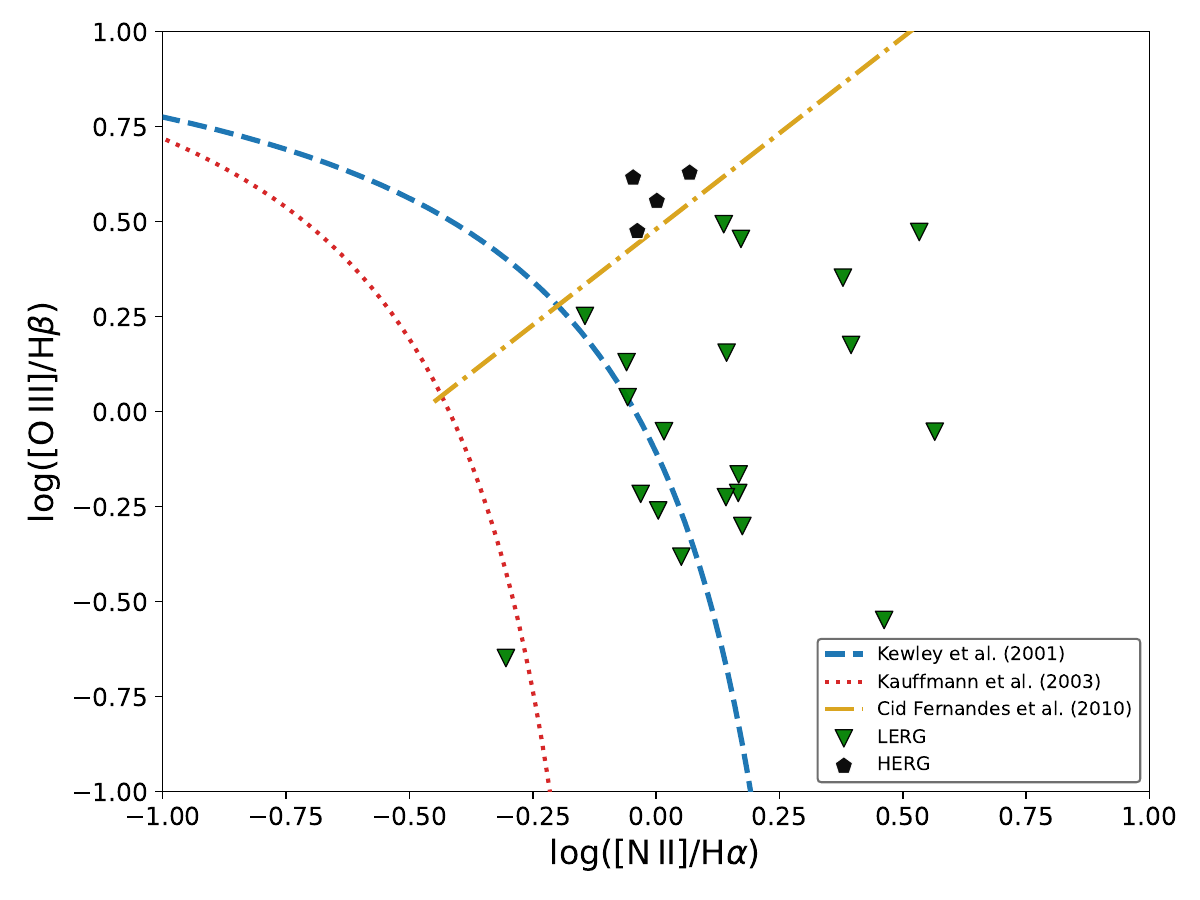}
    \caption{BPT diagnostic diagram showing the emission-line ratios 
    $\log([\mathrm{N\,II}]/\mathrm{H}\alpha)$ versus $\log([\mathrm{O\,III}]/\mathrm{H}\beta)$ for the SDSS-selected sample. 
    The demarcation curves from \citet{kauffmann03apj}, \citet{kewley2001apj}, and \citet{cid10mnras} are shown to separate star-forming, composite, LERG, and HERG regimes. 
    Sources located below the orange dashed–dotted line are classified as LERGs (green triangles), while those above are classified as HERGs (black pentagons).}
    \label{fig:bpt}
\end{figure}

We then cross-checked the Sloan Digital Sky Survey (SDSS\footnote{SDSS website: \url{www.sdss.org}}) for optical spectra for the remaining sources and identified 24 objects with sufficient emission-line information to be classified using the BPT diagram, following the procedure of \citet{chilufya25mnras}. 
Figure~\ref{fig:bpt} shows the distribution of the SDSS subsample in the BPT diagram, with the delimitation curves from \citet{kauffmann03apj}, \citet{kewley2001apj}, and \citet{cid10mnras} separating the star-forming, composite, LERG, and HERG regimes. 
Among the SDSS sources with measured $[\mathrm{N\,II}]/\mathrm{H}\alpha$ and $[\mathrm{O\,III}]/\mathrm{H}\beta$ ratios, 20 lie below the orange dashed–dotted line and are classified as LERGs (green triangles), while 4 fall above the boundary and are classified as HERGs (black pentagons). 
In total, the BPT analysis yields 36 LERGs and 11 HERGs.

For unclassified radio galaxies, we assigned tentative classifications based on partial spectral information. 
Sources exhibiting prominent broad emission lines (e.g., 3C~120) are classified as HERGs, whereas those displaying very weak or absent emission lines, consistent with typical LERG spectra in  \citet{buttiglione10aa}, are classified as LERGs. 
In total, 18 radio galaxies were classified in this case, yielding 12 LERGs and 6 HERGs, as shown in Table~\ref{tab:tentative}.

\begin{table*}[bpht]
\setlength{\tabcolsep}{5pt} 
\tiny
\centering
\caption{Parameters for \textit{Fermi} radio galaxies with the classification based on the partial spectral information.}\label{tab:tentative}
\begin{tabular}{ccccccccccccccc}
\hline
\hline 
 {4FGL Name} &Other name&  $z$ & {Class} & Ref  & $\Gamma_{\gamma}$ &  {$\log L_{\gamma}$} &  {CD}  & Ref & $\log L_{\rm{bol}}$ & Ref & $\log (L_{\rm{bol}}/L_{\rm{Edd}})$ &Ref\\\
 (1)&(2)&(3)& (4) & (5) &(6) & (7) &(8) & (9)&(10) &(11)&(12)&(13)\\
\hline
J0009.7-3217	&	IC 1531	&	0.025	&	LERG	&	B18	&	2.26	&	42.19	&	0.02	&	C23	&	41.91	&	B18	&	-4.54		&	 M14-T02		\\
J0037.9+2612	&	WISE J003719.15+261312.6	&	0.148	&	HERG	&	SDSS	&	2.45	&	43.65	&	0.74	&	P21	&	43.53	&	SDSS	&	-2.79		&	 M14-T02		\\
J0057.7+3023	&	NGC 315	&	0.016	&	LERG	&	H97	&	2.39	&	41.89	&	0.19	&	C23	&	42.97	&	H97	&	-4.06		&	B16		\\
J0312.9+4119	&	B3 0309+411B	&	0.134	&	HERG	&	B89	&	2.47	&	43.61	&	1.35	&	C23	&		&		&			&			\\
J0322.6-3712e	&	Fornax A	&	0.005	&	LERG	&	B92	&	2.07	&	41.31	&	0.20	&	C23	&	40.68	&	K03	&	-5.75		&	M14-T02		\\
J0433.0+0522	&	3C 120	&	0.033	&	HERG	&	S72	&	2.79	&	42.83	&	0.18	&	C23	&	46.10	&	J16	&	0.19		&	H20		\\
J0627.0-3529	&	PKS 0625-35	&	0.054	&	LERG	&	I10	&	1.91	&	43.89	&	0.35	&	C23	&	44.02	&	MI14	&	-3.20		&	MI14		\\
J0858.1+1405	&	3C 212	&	1.048	&	HERG	&	SDSS	&	2.46	&	45.65	&		&		&		&		&			&			\\
J1219.6+0550	&	NGC 4261	&	0.007	&	LERG	&	H97	&	2.17	&	41.11	&		&		&	42.50	&	B09	&	-4.43		&	M14-T02		\\
J1306.7-2148	&	PKS 1304-215	&	0.126	&	LERG	&	G83	&	2.17	&	44.13	&	16.98	&	C23	&		&		&			&			\\
J1435.5+2021	&	TXS 1433+205	&	0.748	&	HERG	&	SDSS	&	2.20	&	45.76	&		&		&		&		&			&			\\
J1630.6+8234	&	NGC 6251	&	0.025	&	LERG	&	W00	&	2.35	&	42.95	&	0.58	&	C23	&	42.90	&	H99	&	-4.07		&	H99		\\
J0840.8+1317	&	3C 207	&	0.681	&	LERG	&	SDSS	&	2.48	&	45.33	&	1.41	&	P21	&		&		&			&			\\
J1843.4-4835	&	PKS 1839-48	&	0.11	&	LERG	&	MI14	&	2.02	&	43.60	&	0.29	&	C23	&	43.51	&	MI14	&	-3.92		&	MI14		\\
J2156.0-6942	&	PKS 2153-69	&	0.028	&	HERG	&	S14	&	2.83	&	41.94	&	0.33	&	C23	&	43.63	&	E10	&	-2.94		&	 M14-T02		\\
J0946.0+4735	&	RX J0946.0+4735	&	0.569	&	LERG	&	SDSS	&	2.03	&	44.85	&		&		&	44.46	&	SDSS	&	-2.42		&	 S-T02		\\
J2329.7-2118	&	PKS 2327-215	&	0.28	&	LERG	&	J09	&	2.31	&	44.80	&	1.66	&	C23	&		&		&			&			\\
J1516.8+2918	&	RGB J1516+293	&	0.13	&	LERG	&	SDSS	&	1.95	&	43.64	&	0.68	&	P21	&	42.65	&	SDSS	&	-4.10		&	 S-T02		\\

\hline
\end{tabular}
\medskip

\small
Notes: Col. (1) the source name; Col. (2) the other name; Col. (3) redshift; Col. (4)-(5) classification and its reference; Col. (6) $\gamma$-ray photon index; Col.(7) $\gamma$-ray luminosities in the unit of erg/s; Col.(8)-(9) the Compton dominance and its reference; Col. (10)-(11) the bolometric luminosity and its reference; Col.(12)-(13) the accretion rate and its reference.
If the stellar velocity dispersions from the SDSS  and/or HyperLeda website (\url{http://atlas.obs-hp.fr/hyperleda}, \citealt{makarov14AA}) are available, then the BH masses are computed from the relation between the BH masses and velocity dispersion \citep{tremaine2002ApJ} (S-T02 or M14-T02); otherwise, the BH masses are obtained from the literature. References: 
S72 for \citet{shields72apj}, 
G83 for \citet{grandi83},
B89 for \citet{bru89}, 
B92 for \citet{baum92}, 
H97 for \citet{ho1997ApJS},  
H99 for \citet{ho1999ApJ}, 
W00 for \citet{werner00}, 
K03 for \citet{kim03apj},     
B09 for \citet{buttiglione09}, 
J09 for \citet{jones09}, 
E10 for \citet{evans10}, 
I10 for \citet{inskip2010MNRAS}, 
S14 for for \citet{sadler2014MNRAS}, 
MI14 for \citet{mingo14mnras}, 
B16 for \citet{bosch16}, 
J16 for \citet{janiak2016MNRAS},  
B18 for \citet{bassi18}, 
H20 for \citet{hlabathe2020mnras}, 
B21 for \citet{balasu2021ApJ},  
P21 for \citet{paliya21apjs}; 
C23 for \citet{chen23apj_sed},
P23 for \citet{pannikkote23apj}.
SDSS for Sloan Digital Sky Survey (\url{www.sdss.org}, \citealt{sdss_cite}).
\end{table*}

\textit{Fermi} observations have also established the presence of $\gamma$-ray FR~0s. 
The first source, Tol~1326$-$379, was identified by \citet{grandi16mnras} and subsequently examined by \citet{fu23raa}. 
Stacking analyses by \citet{paliya21apjl} revealed three additional FR~0 candidates, and \citet{pannikkote23apj} expanded this number to 7, all classified as LERGs based on the BPT diagram \citep{grandi16mnras,pannikkote23apj}. 
Since LEDA~55267 is already included in \textit{Fermi} catalogue, we incorporate the seven FR~0 LERGs [Col. (6) in Table \ref{tab:2}] to further enlarge our \textit{Fermi} radio galaxy sample.

In summary, our final \textit{Fermi} radio galaxy sample comprises 55 LERGs and 17 HERGs. The LERGs span a redshift range of $0.001 \leq z \leq 0.681$ with an average value of $\langle z_{\rm{L}} \rangle = 0.11 \pm 0.02$, while the HERGs span $0.028 \leq z \leq 1.048$ with an average value of $\langle z_{\rm{H}} \rangle = 0.23 \pm 0.07$.

In the $\Gamma_{\gamma}$--$L_{\gamma}$ plane (Fig.~\ref{lum-index}), LERGs tend to occupy the region of lower $\gamma$-ray luminosities ($\langle \log L_{\gamma} \rangle = 43.13 \pm 0.14$ erg/s) and flatter photon indices ($ \langle \Gamma_{\gamma} \rangle = 2.19 \pm 0.04 $), whereas HERGs cluster at higher $\gamma$-ray luminosities ($ \langle \log L_{\gamma} \rangle = 43.93 \pm 0.25 $ erg/s) with steeper photon indices ($\langle \Gamma_{\gamma} \rangle = 2.46 \pm 0.06 $). 
Overall, LERGs and HERGs populate distinct regions in $\Gamma_\gamma-L_{\gamma}$ diagram, similar to the case of blazars \citep{ghisellini11mnras}, supporting a unified scenario in which BL~Lacs correspond to the LERG population and FSRQs align with HERGs.

\section{Discussion}\label{dis}
The physical processes underlying the $\gamma$-rays emission in RLAGNs are not firmly established; however, it is commonly assumed that misaligned AGNs experience the same emission processes as their beamed counterparts (e.g. blazars; \citealt{abdo10b,grandi12ijmps,angioni17}). The relation between the $\gamma$-ray luminosities and the radio-core luminosities was presented in \citet{ghisellini2005aa} for 3 EGRET\footnote{The Energetic Gamma-Ray Experiment Telescope} $\gamma$-ray FR Is (Cen A, NGC 6251 and M 87), which was later confirmed by \citet{grandi12ijmps} with a sample of 11 \textit{Fermi} misaligned RLAGNs (8 FR Is, 3 FR IIs). A similar result was also presented in \citet{maruo14apj} for 12 \textit{Fermi} radio galaxies (8 FR Is, 4 FR IIs) and also FR 0s \citep{khatiya2024apj}, and they found that the slope coefficient of the correlation for radio galaxies is similar to the correlation for blazars. This might indicate that the $\gamma$-ray emission mechanism is similar between radio galaxies and blazars.

However, one must also remember redshift as a significant observational bias. HERGs ($\langle z_{\rm{H}} \rangle = 0.23$) have a higher average redshift than LERGs ($\langle z_{\rm{L}} \rangle = 0.11$). FSRQs are generally observed at higher redshifts ($\langle z_{\rm{F}} \rangle = 1.20$), while BL Lacs are typically found at lower redshifts ($\langle z_{\rm{B}} \rangle = 0.43$). Unfortunately, the definition of BL Lacs ($\text{EW}<5\mathring{A}$; \citealt{stickel.1991.apj.374}) prevents one from getting spectroscopic redshift measurements for BL Lacs if the jet continuum is boosted and significantly dominates the emission-line width \citep{foschini12raa,chen25}. Therefore, an observational bias of redshift could artificially produce the positive tendency from BL Lacs (LERGs) to FSRQs (HERGs) in the plot of $\Gamma_{\gamma}$–$L_{\gamma}$, and obscure the true physical origin of the separation. To mitigate the redshift bias, we adopted the Compton dominance (CD) parameter, defined as the ratio of the flux densities of the IC to the synchrotron peak emission, and other statistical tests to explore the redshift dependence.

\subsection{Compton Dominances and unifications}
The CD is essentially a redshift-independent quantity, defined as the ratio of the IC peak luminosity ($L_{\rm{IC}}$) to the synchrotron peak luminosity ($L_{\rm{syn}}$) or the ratio of the IC peak flux ($f_{\rm{IC}}$) to the synchrotron peak flux ($f_{\rm{syn}}$), ${\rm CD} = \nu_{\rm{IC}} L_{\rm{IC}}/{\nu_{\rm{syn}} L_{\rm{syn}}}=\nu_{\rm{IC}} f_{\rm{IC}}/{\nu_{\rm{syn}} f_{\rm{syn}}}$. The CD is a high-energy to low-energy emission ratio, reducing the redshift dependence \citep{abdo10sed,paliya21apjs}.

\begin{figure}[bpht]
    \centering
    \includegraphics[width=1.0\linewidth]{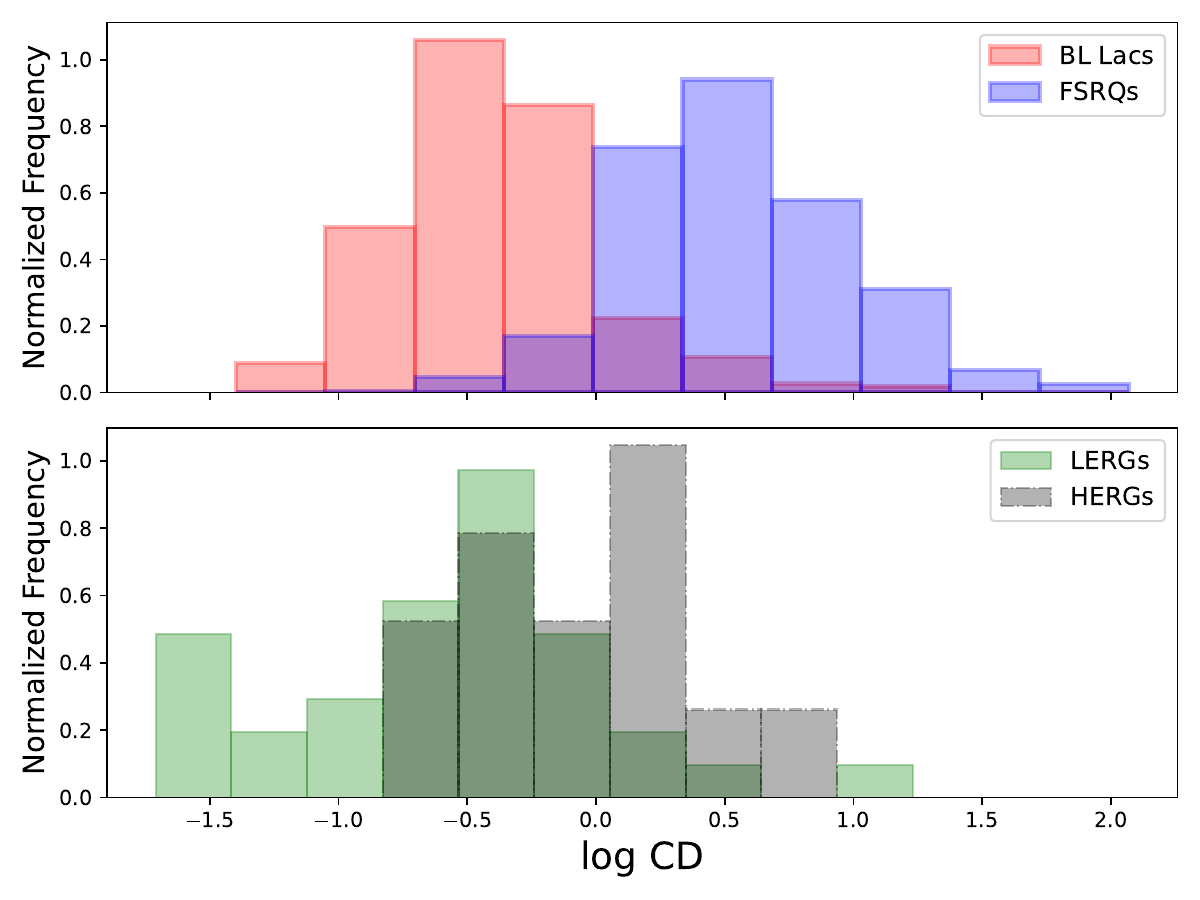}
    \caption{The distribution of the Compton dominances (CD) for blazars (upper panel) and radio galaxies (lower panel). Both of them scale into the same bin.}
    \label{cd_dis}
\end{figure}

We compiled CD values for \textit{Fermi} blazars and radio galaxies from \citet{paliya21apjs} and \citet{chen23apj_sed}. In total, CD values were obtained for 882 blazars and 48 radio galaxies. The resulting CD distributions are shown in Fig.~\ref{cd_dis}. A K-S test comparing the CD distributions of 324 BL Lacs and 558 FSRQs yields a probability of $p<10^{-4}$, indicating that the two populations are statistically distinct. As discussed in \citet{paliya21apjs}, FSRQs typically exhibit higher CD values and are thus more strongly Compton-dominated, whereas BL Lacs tend to have lower CD values, reflecting a weaker Compton component. The dividing line at $\mathrm{CD} = 1$ corresponds to the transition where the high-energy peak becomes dominant over the synchrotron peak, marking a shift in the underlying leptonic emission mechanism from SSC–dominated to EC-dominated scenarios.

For SSC-dominated sources, the SSC emission is related to $\gamma$-ray Doppler factors as: $f_{\rm{SSC}}\sim\delta_{\rm{\gamma}}^{3+\alpha_{\gamma}}$, 
and the synchrotron emission is related to the radio Doppler factors: $f_{\rm{syn}}\sim\delta_{\rm{\rm{r}}}^{3+\alpha_{\rm{r}}}$. The radio spectral index for blazars is flat with the assumption of $\alpha_{\rm{r}}\sim0$ \citep{capetti02aa,abdo10sed,fan16}, and the $\gamma$-ray Doppler factors from \citet{zhang02pasj}, \citet{zhang20apj} and \citet{chen24apjs} are in the similar ranges and averages to the variability Doppler factors from the long-term radio observations \citep{liodakis18apj}, namely, $\delta_{\rm{\rm{\gamma}}}\approx\delta_{\rm{\rm{r}}}$. Therefore, the CD for the SSC-dominated sources may serve as: 
\begin{equation}
    \log \rm{CD} = 
    \log\left(\frac{\textit{f}_{\rm{SSC}}}{\textit{f}_{\rm{syn}}}\right) \propto 
    \log \left( \frac{\delta_{\gamma}^{3+\alpha_\gamma}}{\delta_{\rm{r}}^3} \right)
    \propto \log (\delta_{\gamma}) \cdot \alpha_\gamma, 
\end{equation}

\begin{figure}[bpht]
    \centering
    \includegraphics[width=0.95\linewidth]{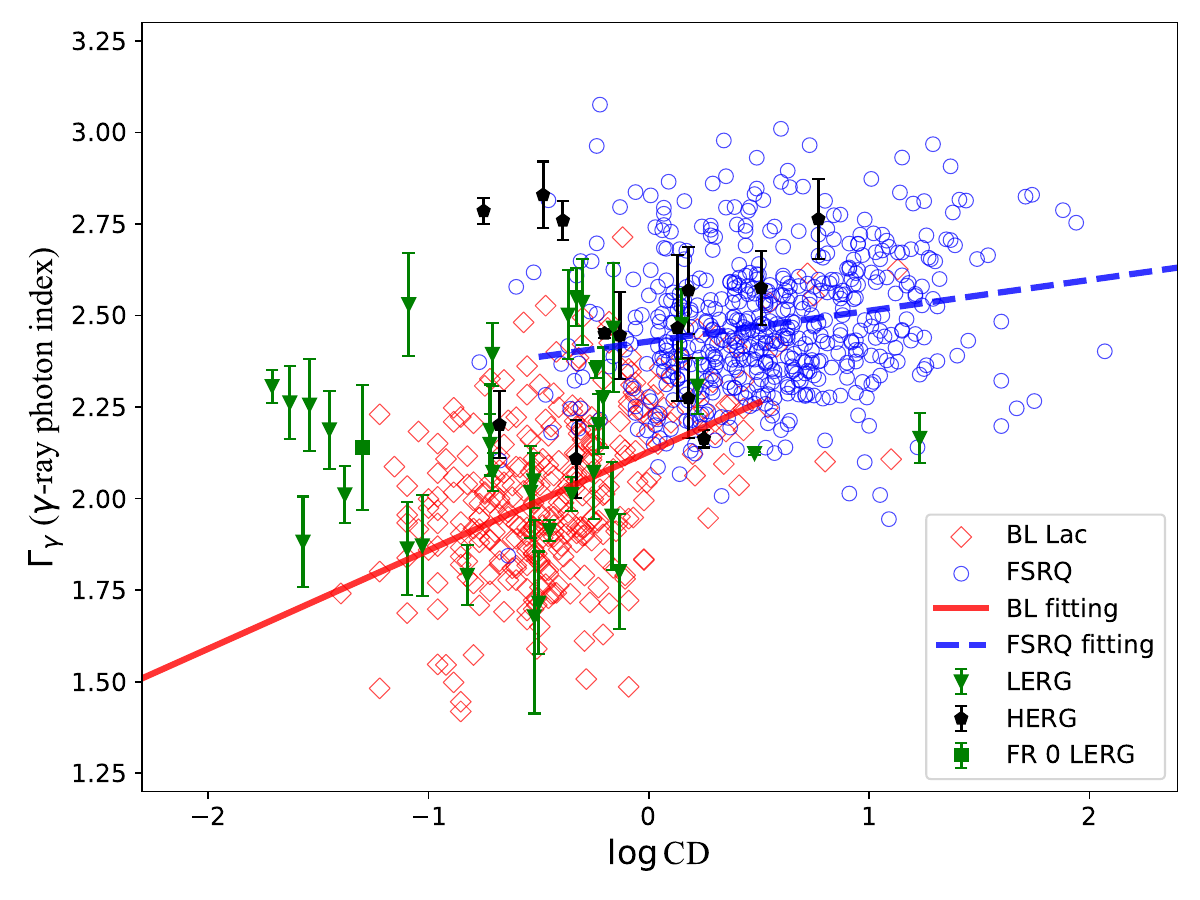}
    \caption{The plot of the relation between the Compton dominance (CD) and the $\gamma$-ray photon indices ($\Gamma_{\gamma}$). The labels are the same as Fig. \ref{lum-index}. The red solid line is the linear regression for  BL Lacs [$\Gamma_{\gamma} =(0.27 \pm 0.03)\,\log \mathrm{CD} + (2.13 \pm 0.02)$], and the blue dashed line is the linear regression for FSRQs [$\Gamma_{\gamma}=(0.08 \pm 0.02)\,\log \mathrm{CD} + (2.43 \pm 0.01)$].  
    }
    \label{cd-index}
\end{figure}

However, for EC-dominated sources, the CD goes with the $\gamma$-ray spectral index as: 
\begin{equation}
    \log \rm{CD} = 
    \log\left(\frac{\textit{f}_{\rm{EC}}}{\textit{f}_{\rm{syn}}}\right) \propto 
    \log \left( \frac{\delta_{\gamma}^{4+2\alpha_\gamma}}{\delta_{\rm{r}}^3} \right)
    \propto (2\log \delta_{\gamma}) \cdot \alpha_\gamma , 
\end{equation}
Therefore, in the plot of the $\gamma$-ray spectral index and the CD ($\Gamma_{\gamma}-$CD), we would expect a broken slope changing from SSC-dominated sources (e.g. BL Lacs) to EC-dominated sources (e.g. FSRQs).
The expected broken slope between BL Lacs and FSRQs is confirmed in Fig. \ref{cd-index}. When a linear regression in $\Gamma_{\gamma}-$CD diagram is considered between BL Lacs and FSRQs, we obtained,
\[
\Gamma_{\gamma} =
\begin{cases}
(0.27 \pm 0.03)\,\log \mathrm{CD} + (2.13 \pm 0.02), & \text{BL Lacs,\ }\\
(0.08 \pm 0.02)\,\log \mathrm{CD} + (2.43 \pm 0.01), & \text{FSRQs,}
\end{cases}
\]
where a correlation coefficient of 0.44 and a probability of $p<10^{-4}$ for BL Lacs, while a correlation coefficient of 0.21 and a probability of $p<10^{-4}$ for FSRQs. The corresponding linear regressions for both BL Lacs (red solid line) and FSRQs (blue dashed line) are shown in Fig. \ref{cd-index}.

As can be seen in the lower panel of  Fig. \ref{cd_dis}, a potential bimodality also appears with 35 LERGs having an average $\langle \rm{logCD_{L}}\rangle = -0.03\pm0.11$, whereas 13 HERGs have a higher average $\langle \rm{logCD_{H}}\rangle = 0.15\pm0.13$,  supporting that an EC mechanism starts dominating in HERGs. The K-S test reveals differences in CD between 35 LERGs and 13 HERGs with $p=0.05$. 
A potential bimodality of CD between LERGs and HERGs is similar to blazars between BL Lacs and FSRQs \citep{paliya21apjs}.

In the plot of $\Gamma_\gamma-$CD, we could find that LERGs are located with the BL Lacs, with the smaller CD and harder spectral index, while HERGs are together with the FSRQs, sharing the higher CD and softer spectral index
These results invoke common signatures of jet radiation mechanisms between the two classes of blazars and radio galaxies, in which BL Lacs reconcile with LERGs with the SSC-dominated radiation, whereas FSRQs reconcile with HERGs with the EC-dominated process.

In summary,  
the $\Gamma_{\gamma}$–CD plane is able to simultaneously
distinguish BL Lacs from FSRQs, as well as LERGs from HERGs, strengthening the connection between the two classes of blazars and radio galaxies through common accretion-ejection physics.

\subsection{Accretion and ejection}\label{acc_eje}

The dichotomy observed in the $\Gamma_{\gamma}-L_{\gamma}$ plane (Fig.~\ref{lum-index}) emerges for both blazars and their misaligned counterparts. On average, HERGs exhibit higher $\gamma$-ray luminosities and softer photon indices than LERGs. In HERGs, the external low-energy seed photons from the broad line region, the torus and the radiatively efficient disc (present in HERGs and not in LERGs, \citealt{Tadhunter08,heckman2014araa}) are efficiently up-scattered by relativistic jets, leading to the higher $\gamma$-ray luminosities with softer $\gamma$-ray spectra due to the faster energy losses of the relativistic electrons. In contrast, the SSC process in LERGs relies on seed photons originating from the internal synchrotron emission of the jets, producing the relatively lower $\gamma$-ray luminosities with the harder $\gamma$-ray spectra observed by \textit{Fermi}. This result supports a unified picture: the $\gamma$-ray emissions of both BL Lacs and LERGs are powered by SSC, while those of FSRQs and HERGs are dominated by EC.

While this framework primarily applies to $\gamma$-ray emission produced in the inner jet regions, additional contributions from external seed photons on larger spatial scales may also be relevant in some sources. A notable example is the nearby radio galaxy Centaurus~A, for which extended $\gamma$-ray emission has been detected from its radio lobes \citep{abdo_sci_2010_cena}. In this case, the lobe component accounts for more than half of the total $\gamma$-ray output and is commonly interpreted as IC scattering of CMB photons. This indicates that, in \textit{Fermi} radio galaxies with significant extended emission, the observed $\gamma$-ray luminosity may not be directly linked to the central engine or to the accretion disc properties.

\begin{figure}
    \centering
    \includegraphics[width=1.0\linewidth]{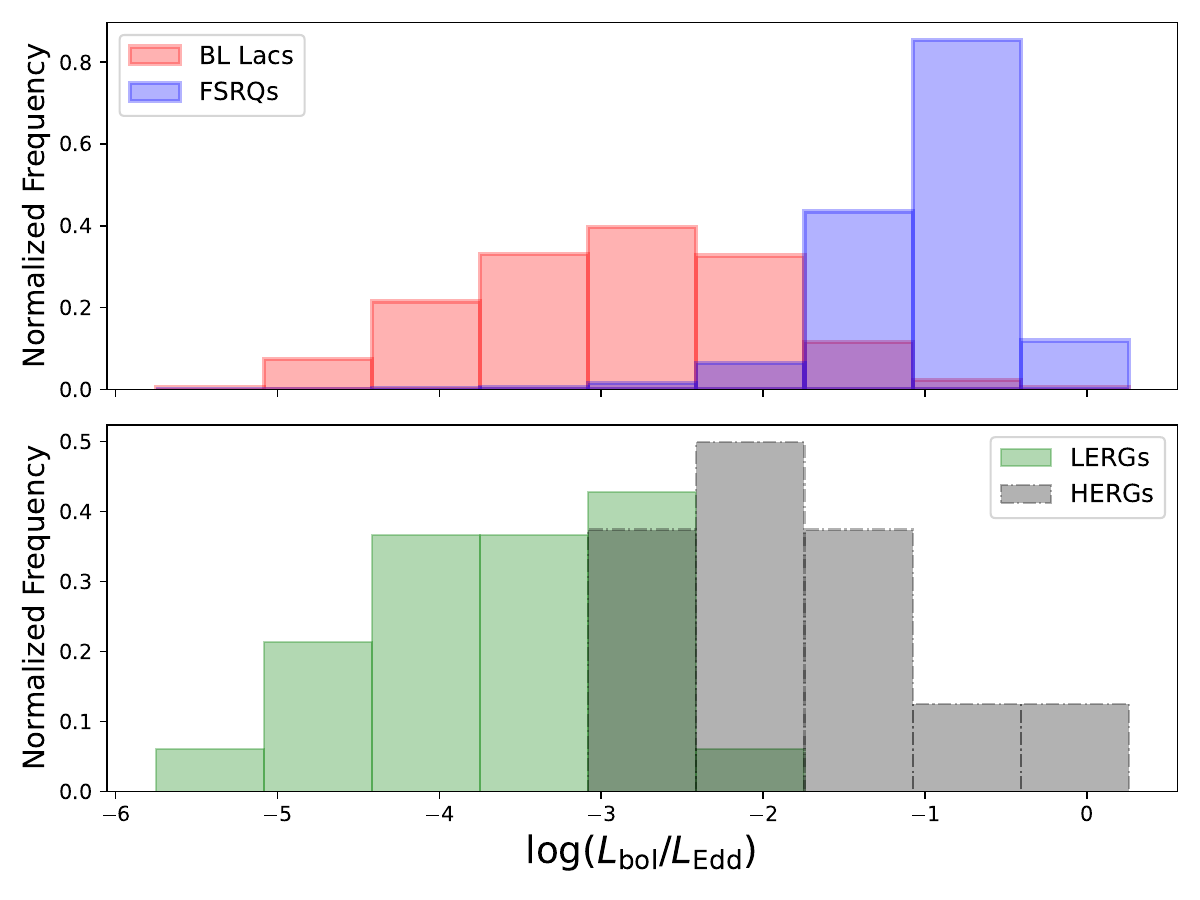}
    \caption{The distribution of the accretion rates ($L_{\rm{bol}}/L_{\rm{Edd}}$) for blazars (upper panel) and radio galaxies (lower panel). Both of them scale into the same bin.}
    \label{acc_dis}
\end{figure}

Furthermore, \citet{paliya21apjs} recenlty promoted that the CD can be considered as a powerful parameter to reveal the physics of the central engine in beamed AGNs: the low-accretion-rate sources have lower CD, and high-accretion-rate sources exhibit larger CD. This positive tendency is later discussed and supported with a large \textit{Fermi} sample including \textit{Fermi} blazars, narrow-line Seyfert galaxies, and radio galaxies \citep{chen23apj_sed}. These results strengthen the argument that CD can be considered a good proxy for the accretion rate in RLAGNs. For further support of the accretion rate differences,  we also collected the bolometric (thermal) luminosities and the BH masses from \citet{paliya21apjs}, in which they compiled a sample of \textit{Fermi} blazars with the spectroscopic information from SDSS, and obtained the bolometric luminosities from the broad emission-line luminosities with the assumption, $L_{\rm{BLR}}=0.1 L_{\rm{bol}}$ \citep{ghi14nature}. However, for BL Lacs with featureless emission spectra, the bolometric luminosities are estimated from the $3\sigma$ upper limits of the total emission-line luminosity \citep{paliya21apjs}. With the goal of deriving the accretion rate ($L_{\rm{bol}}/L_{\rm{Edd}}$), we collected the BH masses and bolometric luminosities for both BL Lacs and FSRQs from \citet{paliya21apjs} and \citet{chen23apj_sed}. 
The accretion-rate distributions of 320 BL Lacs and 564 FSRQs (Fig.~\ref{acc_dis}) differ significantly, as confirmed by a K–S test with the probability of $p<10^{-4}$.

The accretion rate $L_{\rm{bol}}/L_{\rm{Edd}}$ (i.e. Eddington ratio), for radio galaxies is derived from the bolometric (thermal) emission, estimated from the isotropic [O~III] ($\lambda5007\mathring{A}$) emission line luminosity from the narrow line region, $L_{\rm{bol}}=3500 L_{\rm{[O\ III]}}$ \citep{heckman04apj,heckman2014araa}. If the [O III] emission lines are not available from the literature or SDSS, we considered the Chandra 2-7 keV X-ray luminosity \citep{evans10} as an indicator for the bolometric luminosity \citep{lussomn12,lopez24aa}. We adopted the relation, $L_{\rm{bol}} = 9.588  L_{\rm{X}}$, for LERGs \citep{lopez24aa}, or computed the bolometric luminosity for BLRGs (Type-I) or HERGs (Type-II) from the cubic functions between the bolometric and 2-10 keV X-ray luminosities of \citet{lussomn12}. 

If the stellar velocity dispersions from the SDSS  and/or HyperLeda website (\url{http://atlas.obs-hp.fr/hyperleda}, \citealt{makarov14AA}) are available, then the BH masses are computed from the relation between the BH masses and velocity dispersion \citep{tremaine2002ApJ} (labelled as S-T02 or M14-T02 in the Table \ref{tab:2} and Table \ref{tab:tentative}); otherwise, the BH masses are obtained from the literature. The $\gamma$-ray properties, bolometric luminosities and BH masses for those 7 FR 0 LERGs are directly collected from the \citet{grandi16mnras}, \citet{baldi.18.aa.fr0}, and \citet{pannikkote23apj}. 

Finally, we obtained a sample of 61 radio galaxies (49 LERGs, 12 HERGs) with available accretion rates, as presented in Tab. \ref{tab:2} and Tab. \ref{tab:tentative}. A double peak of $\log (L_{\rm{bol}}/L_{\rm{Edd}})$ between LERGs and HERGs is also shown in the lower panel of Fig. \ref{acc_dis}. The Eddington ratio for LERGs ranges from $\log (L_{\rm{bol}}/L_{\rm{Edd}})=-5.75$ to  $\log (L_{\rm{bol}}/L_{\rm{Edd}})=-1.88$, with an average of $-3.57\pm0.12$, while for HERGs, it ranges from $\log (L_{\rm{bol}}/L_{\rm{Edd}})=-2.94$ to $\log (L_{\rm{bol}}/L_{\rm{Edd}})=0.19$, with an average of $-1.79\pm0.26$.

A K-S test probability for the Eddington ratio between 49 LERGs and 12 HERGs is $p<10^{-4}$, supporting that their accretion rates are significantly different, as also reported from other studies (e.g. \citealt{best12mnras,grandi21,chilufya25mnras,arnaudova25}).  
This bimodality not only implies a switch of the disc from inefficient to efficient, but also towards a unified accretion scenario between radio galaxies and blazars: low efficient disc (e.g. ADAF; \citealt{heckman2014araa,torresi22MnSAI,ye25aa}) in both LERGs and BL Lacs, and high efficient standard disc in both HERGs and FSRQs.

\begin{figure}
    \centering
    \includegraphics[width=0.99\linewidth]{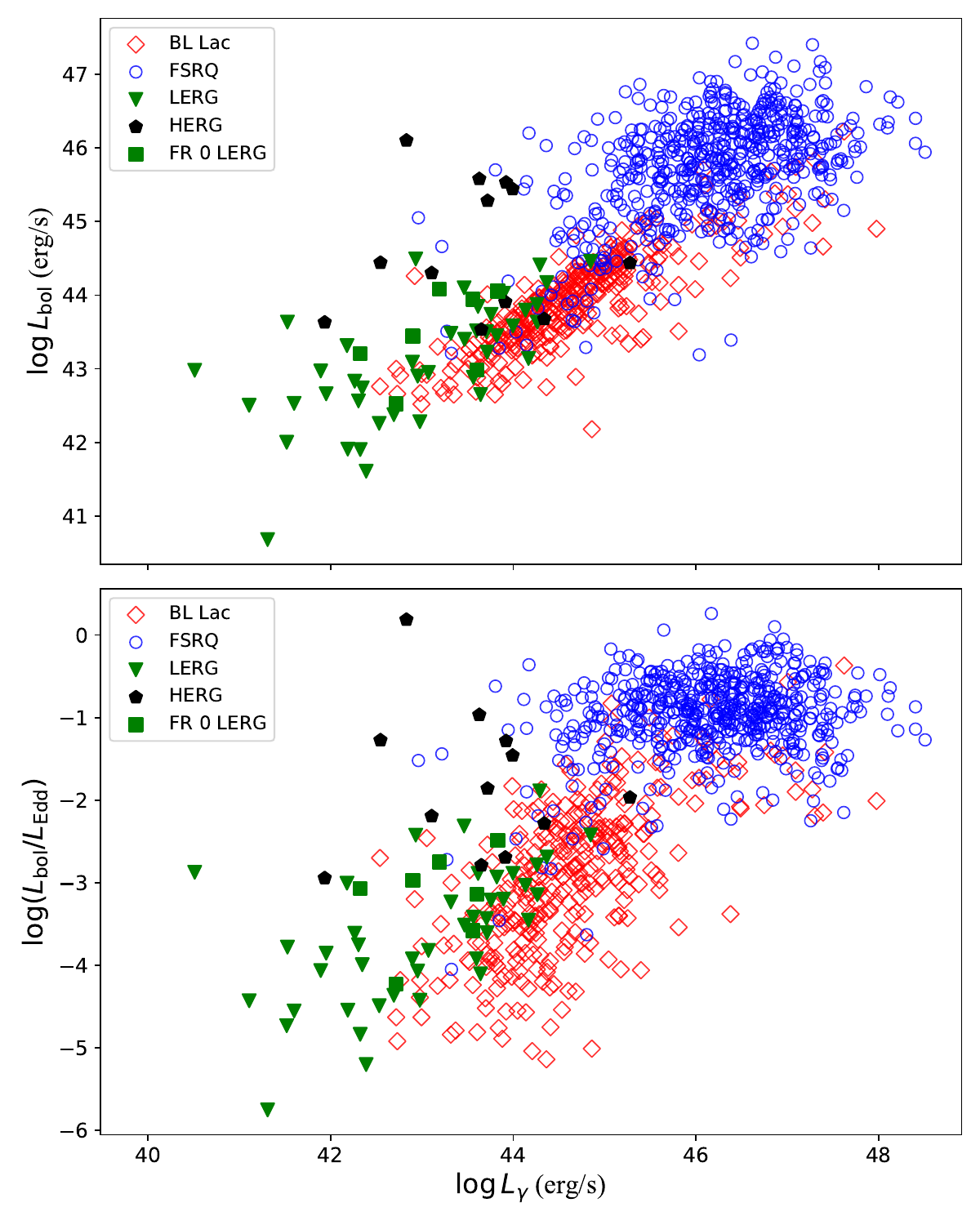}
    \caption{The relation between the $\gamma$-ray luminosities $L_{\gamma}$ and the bolometric luminosities $L_{\rm{bol}}$ (upper panel), or the accretion rate $L_{\rm{bol}}/L_{\rm{Edd}}$ (lower panel) for blazars and radio galaxies. The labels are the same as Fig. \ref{lum-index}.}
    \label{fig:lum_disc_acc}
\end{figure}

The symbiotic relationship between accretion and ejection in blazar and radio galaxies is strongly supported by population studies \citep{ghisellini01aa,abdo.2010.apj.720.misagn,xie12pasj,sbarrato12mnras,sbarrato14mnras,ghi14nature,chen15aj,paliya21apjs}. The relation between the bolometric luminosities (indicator for accretion) and the $\gamma$-ray luminosities (indicator for jet) is plotted in Fig. \ref{fig:lum_disc_acc}. The strong correlation between them is evident for both blazars and radio galaxies, as shown in the upper panel of Fig. \ref{fig:lum_disc_acc}. Whereas for HERGs the statistics are limited by the small number of objects, these tight relationships persist for BL Lacs, FSRQs, and LERGs even after applying the Spearman partial rank correlation\footnote{The Spearman partial rank correlation coefficient estimates the correlation between two variables while accounting for the influence of a third. If variables A and B are both correlated with variable $z$, the partial correlation is given by $\rho_{{\rm AB,}z}=\frac{\rho_{\rm AB}-\rho_{{\rm A}z}\rho_{{\rm B}z}}{\sqrt{(1-\rho^2_{{\rm A}z})(1-\rho^2_{{\rm B}z})}}$.}, which considers the effect of redshift (see Table \ref{partial}), with probability of $p<10^{-4}$. In addition, we can note that LERGs and BL Lacs show parallel relations, with the latter, of course, $\gamma$-ray-louder. These results further support the existence of accretion–jet symbiosis in RLAGNs.

\begin{table*}[bpht]
\centering
\caption{Spearman partial rank correlation results for $L_{\gamma}$-$L_{\rm bol}$ for {\it Fermi} RLAGN (Fig.~\ref{fig:lum_disc_acc}).}
\label{partial}
\begin{tabular}{lccc}
\hline\hline
Class & Spearman $\rho_{\rm AB}$ & Probability & Partial Spearman $\rho_{{\rm AB},z}$ \\
\hline
FSRQ    & 0.52  & $<10^{-4}$ & 0.05 \\
BL Lac  & 0.83  & $<10^{-4}$ & 0.35 \\
LERG    & 0.66 & $<10^{-4}$ & 0.48 \\
HERG    & 0.04  & $0.91$ & 0.16 \\
\hline
\end{tabular}
\vspace{0.5em}
\begin{minipage}{0.85\textwidth}
\small
\textbf{Note.} $\rho_{\rm AB}$ is the standard Spearman correlation coefficient, and the corresponding probability refers to the null hypothesis of no correlation. $\rho_{{\rm AB},z}$ is the Spearman partial rank correlation coefficient, which accounts for the common dependence of variables A ($\gamma$-ray luminosity) and B (bolometric luminosity) on $z$ (redshift).
\end{minipage}
\end{table*}

In the lower panel of Fig.~\ref{fig:lum_disc_acc}, we presented the relation between $\gamma$-ray luminosity and accretion rate [$\log (L_{\rm{bol}}/L_{\rm{Edd}})$] for RLAGNs. Notably, the slope changes significantly from BL Lacs to FSRQs, implying the presence of two distinct central engines separated at $\log (L_{\rm{bol}}/L_{\rm{Edd}}) \sim -2$ (also evident in Fig.~\ref{acc_dis}), which is the standard separation from (quasar-like) radiatively efficient and (ADAF-type) radiatively-inefficient disc physics \citep{ghisellini11mnras,heckman2014araa}. Interestingly, for BL Lacs, $\gamma$-ray luminosity increases with the Eddington ratio, whereas in FSRQs, the relation between the accretion rate and $\gamma$-ray luminosity becomes nearly flat above $\log (L_{\rm{bol}}/L_{\rm{Edd}}) > -2$. 
However, the contribution of Doppler boosting to the $\gamma$-ray luminosities and the dependence on the jet orientation and bulk velocity make this diagram difficult to interpret in relation to two accretion and ejection states. Further studies on jet and high-energy emission for RLAGNs could provide new light to understand the physical segregation of different classes noted in Fig. \ref{fig:lum_disc_acc}.


In comparison, \textit{Fermi} radio galaxies, with accretion rates comparable to those of blazars, exhibit systematically lower $\gamma$-ray luminosities, forming a roughly parallel correlation. Assuming a linear relation between the bolometric luminosity (or accretion rate) and the $\gamma$-ray luminosity \citep{ghi14nature}, i.e. $\log L_{\rm bol} = a \log L_\gamma^{\rm ob} + b$, and adopting $\log L_\gamma^{\rm ob} = \log L_\gamma^{\rm in} + p \log \delta$ when the observed $\gamma$-ray emission is Doppler-boosted for blazars, we obtain $\log L_{\rm{bol}}= a\times\log L^{in}_\gamma + [b+(a*p)\log\delta]$, where $a$, $b$, and $p$ are the slope, intercept, and the different jet models\footnote{$p = 2 + \alpha$ is for a continuous jet model, and  $p = 3 + \alpha$ is for a spherical jet model \citep{ghisellini.1993.apj.407,urry.1995.pasp.107}.} respectively. Since the bolometric luminosity (and hence accretion rate, because it is derived from the emission lines) is an isotropic quantity for both radio galaxies and blazars, the slopes of the two populations are expected to be similar. The parallel relations observed in Fig.~\ref{fig:lum_disc_acc} support a unified accretion–ejection scenario, at least, for LERGs and BL Lacs. Owing to the limited HERG sample, no FSRQ-like trend is identifiable among the HERGs.

\begin{figure}
    \centering
    \includegraphics[width=1.0\linewidth]{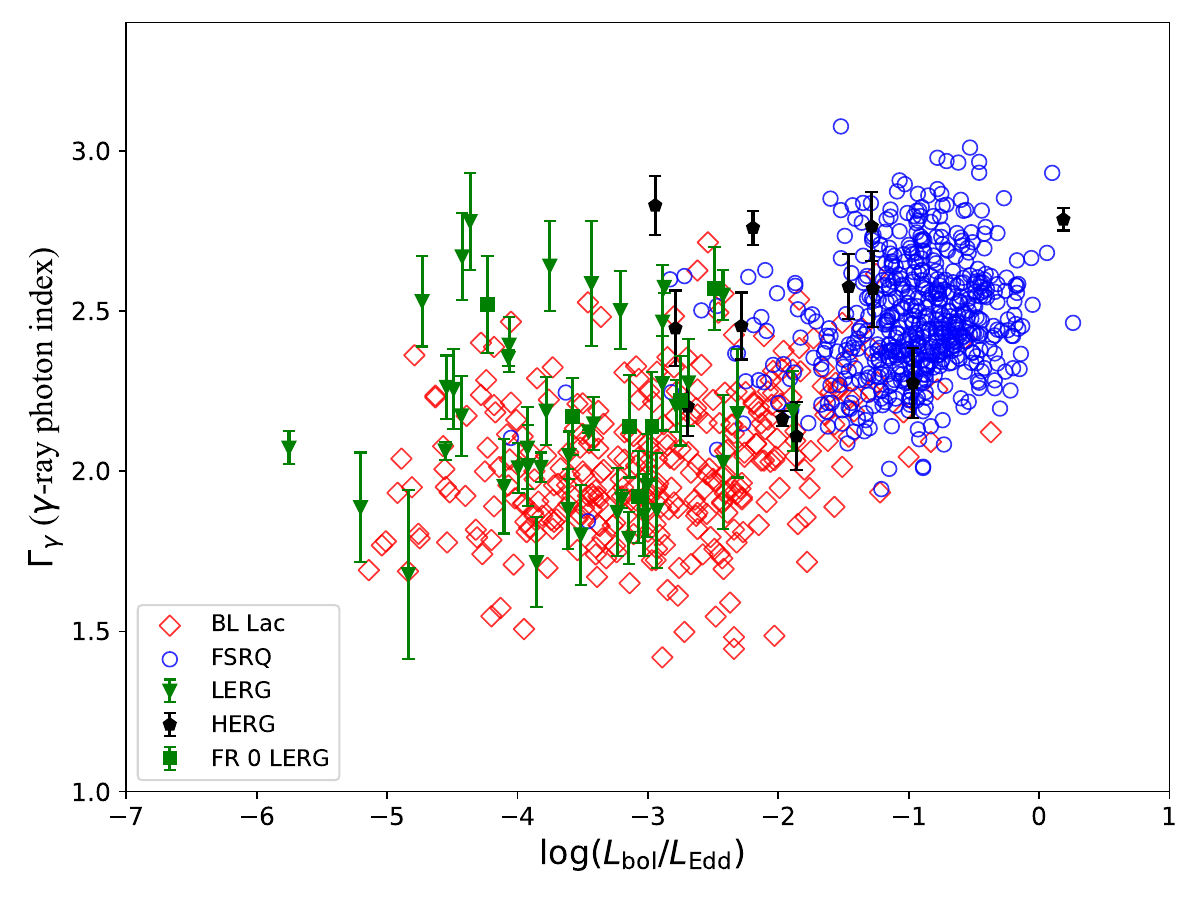}
    \caption{The plot of the relation between the accretion rates ($L_{\rm{bol}}/L_{\rm{Edd}}$) and $\gamma$-ray photon indices ($\Gamma_\gamma$). The labels are the same as Fig. \ref{lum-index}. }
    \label{acc-index}
\end{figure}

This observed bimodality in the accretion and ejection is also clearly reflected in the $\Gamma_{\gamma}$–$L_{\rm{bol}}/L_{\rm{Edd}}$ plane (Fig.~\ref{acc-index}), where a noticeable change in slope is observed from BL Lacs to FSRQs. This pronounced transition strongly supports the shift between SSC- and EC-dominated emission mechanisms. At low accretion rates ($\log (L_{\rm{bol}}/L_{\rm{Edd}}) < -2$), the $\gamma$-ray spectral index remains, more or less, flat for BL lacs. However, as the accretion rate becomes higher ($\log (L_{\rm{bol}}/L_{\rm{Edd}}) > -2$) to form a standard disc, the spectral index becomes significantly softer for FSRQs. Within a high-radiative disc environment, the presence of abundant external seed photons in the broad line region or dusty torus, stimulated by the central accretion disc, effectively cools the relativistic electrons, leading to a rapid steepening of the $\gamma$-ray spectrum. This provides clear evidence for the accretion–ejection connection and the existence of distinct beaming patterns between BL Lacs and FSRQs.  Both the CD and broad emission-line luminosities are anti-correlated with the synchrotron peak frequency, showing a sequence from FSRQs to BL Lacs \citep{paliya21apjs,chen24apjs}. These results suggest the Compton cooling effects as an explanation for the blazar sequence, and also strongly support the accretion-ejection scenario between BL Lacs and FSRQs.

To further test whether the observed differences in blazars could be affected by redshift-related selection bias, we repeated the analysis using redshift-matched blazar subsamples ($z \le 1.048$), comparable to the redshift range of the \textit{Fermi} radio galaxy sample. The statistical results remain consistent with those obtained from the full samples. The K-S tests were performed on the distributions of the $\gamma$-ray luminosity, $\gamma$-ray photon index, CD, and the accretion rate, and we found that all comparisons yield $p < 0.05$, indicating that BL Lacs and FSRQs are significantly different in these properties. Specifically, BL Lacs continue to populate the low-luminosity ($\langle\log L_{\gamma}\rangle=44.78\pm0.03$ erg/s), low spectral-index ($\langle \Gamma_{\gamma}\rangle=2.00\pm0.01$), low Compton-dominance ($\langle\log \rm{CD}\rangle=-0.43\pm0.02$), and low accretion-rate ($\langle\log L_{\rm{bol}}/L_{\rm{Edd}}\rangle=-2.97\pm0.06$) regime, whereas FSRQs are associated with higher luminosities ($\langle\log L_{\gamma}\rangle=45.53\pm0.05$ erg/s), steeper spectral indices ($\langle \Gamma_{\gamma}\rangle=2.43\pm0.01$), higher Compton dominances ($\langle\log \rm{CD}\rangle=0.37\pm0.03$), and higher accretion rates ($\langle\log L_{\rm{bol}}/L_{\rm{Edd}}\rangle=-1.08\pm0.03$). These results confirmed that the trends discussed above are not driven by redshift differences, but instead reflect intrinsic physical distinctions between BL Lacs and FSRQs, in agreement with the full-sample analysis.

 In summary, the accretion rate and  $\gamma$-ray properties well separate the SSC-dominated low accretors (LERGs and BL~Lacs) from the EC-dominated high accretors (HERGs and FSRQs), which supports the scenario of a possible common accretion-ejection scenario between blazars and radio galaxies.

\section{Conclusions}
We compiled a sample of \textit{Fermi} blazars and radio galaxies with the $\gamma$-ray luminosities, $\gamma$-ray spectral indices, CD, and accretion rates to discuss their radiation mechanisms and potential unification between blazars and radio galaxies. Here are the conclusions: 

\begin{itemize}
    \item The distinct regions occupied by BL Lacs and FSRQs in the $\Gamma_\gamma$–$L_\gamma$ plane can be attributed to different beaming patterns associated with SSC and EC mechanisms. We also found a similar tendency from LERGs to HERGs, suggesting common high-energy mechanisms between BL~Lacs-LERGs (SSC) and FSRQs-HERGs (EC).
    
    \item  We analysed the different beaming patterns with the CD, and predicted a slope change from SSC-dominated sources  [$\log \rm{CD}\propto (\log \delta_{\gamma})\cdot\alpha_{\gamma}$]  to EC-dominanted sources [$\log \rm{CD}\propto (2\log \delta_{\gamma})\cdot\alpha_{\gamma}$)]. This phenomenon is confirmed by a sample of 882 blazars. A significant change in the $\Gamma_\gamma$-CD diagram is presented between BL Lacs and FSRQs, in which LERGs and HERGs are also located in a similar area as blazar distributions, better reinforcing the beaming pattern (SSC and EC) models for BL~Lacs-LERGs and FSRQs-HERGs.

    \item  A clear transition from BL Lacs to FSRQs is observed in the $\Gamma_\gamma$–$ L_{\rm{bol}}/L_{\rm{Edd}}$ plane, which clearly provides compelling evidence for the accretion–ejection dichotomy between BL Lacs and FSRQs. Furthermore, the clear differences in blazars also successfully distinguish between LERGs and HERGs, suggesting the existence of two accretion-ejection modes for BL~Lacs-LERGs and FSRQs-HERGs. We also note that the FR~0s are consistent with the LERG population in general \citep{baldi10,baldi19b}.
    
.

\end{itemize}

All these results gather evidence that  BL Lacs-LERGs and FSRQs-HERGs correspond to two different accretion-ejection mechanisms, regardless of the extended radio morphology, which is consistent with the earlier studies \citep{lai94,XU2009apjl,giommi2012mnras}.  In radiatively efficient systems, such as FSRQs and HERGs (generally FR IIs), the accretion disc is capable of illuminating the broad/narrow-line region and dusty torus, creating a photon-rich environment around the relativistic jet. This abundance of external photons leads to efficient cooling of relativistic electrons, preventing them from reaching very high energies. As a result, the high-energy spectral peak occurs at lower frequencies and exhibits a steeper slope in \textit{Fermi} $\gamma$-ray observations. Conversely, in radiatively inefficient accretion systems, such as BL Lacs and LERGs (FR 0s, FR Is, FR IIs), the surrounding nuclear environments remain similar but photon-poor \citep{capetti05}. The scarcity of seed photons allows jet electrons to attain higher energies, and phenomenalizes a high-frequency peak and a flatter $\gamma$-ray spectrum. Further multi-band studies on statistically-complete samples of aligned and misaligned radio galaxies are needed to test the robustness of a unification scheme between FSRQs-HERGs and BL~Lacs-LERGs.

\section*{Data availability}
Full Table \ref{tab:1} is available in electronic form at the CDS via anonymous ftp to \url{cdsarc.u-strasbg.fr (130.79.128.5)} or via \url{http://cdsweb.u-strasbg.fr/cgi-bin/qcat?J/A+A/}.

\begin{acknowledgements}
We greatly appreciate the referee’s support and insightful comments, which have helped us improve the manuscript.
We would also like to appreciate the comments and ideas from E. Torresi, and the discussion with A. Capetti and F. Tavecchio. The work is partially supported by the National Natural Science Foundation of China (NSFC 12433004, U2031201), the Eighteenth Regular Meeting Exchange Project of the Scientific and Technological Cooperation Committee between the People’s Republic of China and the Republic of Bulgaria (Series No. 1802). JHF also acknowledges the science research grants from the China Manned Space Project with NO. CMS-CSST-2025-A07, and the support for Astrophysics Key Subjects of Guangdong Province.  RDB acknowledges financial support from INAF mini-grant \textit{\lq\lq FR0 radio galaxies\rq\rq} (Bando Ricerca Fondamentale INAF 2022).  This research was also partially supported by the Bulgarian National Science Fund of the Ministry of Education and Science under grants KP-06-H68/4 (2022), KP-06-H88/4 (2024), and KP-06-KITAJ/12 (2024). This research has made use of SDSS-III, which has been provided by the Alfred P. Sloan Foundation, the Participating Institutions, the National Science Foundation, and the U.S. Department of Energy Office of Science.  This research has made use of the NASA/IPAC Extragalactic Database (NED), which is operated by the Jet Propulsion Laboratory, California Institute of Technology, under contract with the National Aeronautics and Space Administration. XUHONG YE and WENXIN YANG both acknowledge the support from the Chinese Scholarship Council.
\end{acknowledgements}



\bibliographystyle{aa}
\bibliography{yebib}

@ARTICLE{bru89,
       author = {{de Bruyn}, A.~G.},
        title = "{0309+411, an Mpc-sized core-dominated radio galaxy/quasar.}",
      journal = {\aap},
         year = 1989,
        month = dec,
       volume = {226},
        pages = {L13-L16},
       adsurl = {https://ui.adsabs.harvard.edu/abs/1989A&A...226L..13D}
}

@ARTICLE{m09,
       author = {{Masetti}, N. and {Parisi}, P. and {Palazzi}, E. and {Jim{\'e}nez-Bail{\'o}n}, E. and {Morelli}, L. and {Chavushyan}, V. and {Mason}, E. and {McBride}, V.~A. and {Bassani}, L. and {Bazzano}, A. and {Bird}, A.~J. and {Dean}, A.~J. and {Galaz}, G. and {Gehrels}, N. and {Landi}, R. and {Malizia}, A. and {Minniti}, D. and {Schiavone}, F. and {Stephen}, J.~B. and {Ubertini}, P.},
        title = "{Unveiling the nature of INTEGRAL objects through optical spectroscopy. VII. Identification of 20 Galactic and extragalactic hard X-ray sources}",
      journal = {\aap},
         year = 2009,
        month = feb,
       volume = {495},
       number = {1},
        pages = {121-135},
          doi = {10.1051/0004-6361:200811322},
archivePrefix = {arXiv},
       eprint = {0811.4085},
 primaryClass = {astro-ph},
       adsurl = {https://ui.adsabs.harvard.edu/abs/2009A&A...495..121M}
}

@ARTICLE{b11,
       author = {{Buttiglione}, S. and {Capetti}, A. and {Celotti}, A. and {Axon}, D.~J. and {Chiaberge}, M. and {Macchetto}, F.~D. and {Sparks}, W.~B.},
        title = "{An optical spectroscopic survey of the 3CR sample of radio galaxies with z < 0.3. III. Completing the sample}",
      journal = {\aap},
         year = 2011,
        month = jan,
       volume = {525},
          eid = {A28},
        pages = {A28},
          doi = {10.1051/0004-6361/201015574},
archivePrefix = {arXiv},
       eprint = {1010.1650},
 primaryClass = {astro-ph.CO},
       adsurl = {https://ui.adsabs.harvard.edu/abs/2011A&A...525A..28B}
}

@ARTICLE{chen25,
       author = {{Chen}, Guohai and {Yang}, Wenxin and {Liu}, Yi and {Ho}, Luis C. and {Xiao}, Hubing and {Bachev}, Rumen S. and {Strigachev}, Anton and {Fan}, Junhui},
        title = "{Changing-look behaviour and amplitude-modulated QPO in blazar Ton 599}",
      journal = {\mnras},
         year = 2025,
        month = dec,
       volume = {544},
       number = {2},
        pages = {1926-1938},
          doi = {10.1093/mnras/staf1818},
       adsurl = {https://ui.adsabs.harvard.edu/abs/2025MNRAS.544.1926C}
}

@ARTICLE{abdo_sci_2010_cena,
       author = {{Abdo}, A.~A. and {Ackermann}, M. and {Ajello}, M. and {Atwood}, W.~B. and {Baldini}, L. and {Ballet}, J. and {Barbiellini}, G. and {Bastieri}, D. and {Baughman}, B.~M. and {Bechtol}, K. and {Bellazzini}, R. and {Berenji}, B. and {Blandford}, R.~D. and {Bloom}, E.~D. and {Bonamente}, E. and {Borgland}, A.~W. and {Bregeon}, J. and {Brez}, A. and {Brigida}, M. and {Bruel}, P. and {Burnett}, T.~H. and {Buson}, S. and {Caliandro}, G.~A. and {Cameron}, R.~A. and {Caraveo}, P.~A. and {Casandjian}, J.~M. and {Cavazzuti}, E. and {Cecchi}, C. and {{\c{C}}elik}, {\"O}. and {Chekhtman}, A. and {Cheung}, C.~C. and {Chiang}, J. and {Ciprini}, S. and {Claus}, R. and {Cohen-Tanugi}, J. and {Colafrancesco}, S. and {Cominsky}, L.~R. and {Conrad}, J. and {Costamante}, L. and {Cutini}, S. and {Davis}, D.~S. and {Dermer}, C.~D. and {de Angelis}, A. and {de Palma}, F. and {Digel}, S.~W. and {do Couto e Silva}, E. and {Drell}, P.~S. and {Dubois}, R. and {Dumora}, D. and {Farnier}, C. and {Favuzzi}, C. and {Fegan}, S.~J. and {Finke}, J. and {Focke}, W.~B. and {Fortin}, P. and {Fukazawa}, Y. and {Funk}, S. and {Fusco}, P. and {Gargano}, F. and {Gasparrini}, D. and {Gehrels}, N. and {Georganopoulos}, M. and {Germani}, S. and {Giebels}, B. and {Giglietto}, N. and {Giordano}, F. and {Giroletti}, M. and {Glanzman}, T. and {Godfrey}, G. and {Grenier}, I.~A. and {Grove}, J.~E. and {Guillemot}, L. and {Guiriec}, S. and {Hanabata}, Y. and {Harding}, A.~K. and {Hayashida}, M. and {Hays}, E. and {Hughes}, R.~E. and {Jackson}, M.~S. and {J{\'o}hannesson G.} and {Johnson}, S. and {Johnson}, T.~J. and {Johnson}, W.~N. and {Kamae}, T. and {Katagiri}, H. and {Kataoka}, J. and {Kawai}, N. and {Kerr}, M. and {Kn{\"o}dlseder}, J. and {Kocian}, M.~L. and {Kuss}, M. and {Lande}, J. and {Latronico}, L. and {Lemoine-Goumard}, M. and {Longo}, F. and {Loparco}, F. and {Lott}, B. and {Lovellette}, M.~N. and {Lubrano}, P. and {Madejski}, G.~M. and {Makeev}, A. and {Mazziotta}, M.~N. and {McConville}, W. and {McEnery}, J.~E. and {Meurer}, C. and {Michelson}, P.~F. and {Mitthumsiri}, W. and {Mizuno}, T. and {Moiseev}, A.~A. and {Monte}, C. and {Monzani}, M.~E. and {Morselli}, A. and {Moskalenko}, I.~V. and {Murgia}, S. and {Nolan}, P.~L. and {Norris}, J.~P. and {Nuss}, E. and {Ohsugi}, T. and {Omodei}, N. and {Orlando}, E. and {Ormes}, J.~F. and {Paneque}, D. and {Parent}, D. and {Pelassa}, V. and {Pepe}, M. and {Pesce-Rollins}, M. and {Piron}, F. and {Porter}, T.~A. and {Rain{\`o}}, S. and {Rando}, R. and {Razzano}, M. and {Razzaque}, S. and {Reimer}, A. and {Reimer}, O. and {Reposeur}, T. and {Ritz}, S. and {Rochester}, L.~S. and {Rodriguez}, A.~Y. and {Romani}, R.~W. and {Roth}, M. and {Ryde}, F. and {Sadrozinski}, H.~F.-W. and {Sambruna}, R. and {Sanchez}, D. and {Sander}, A. and {Saz Parkinson}, P.~M. and {Scargle}, J.~D. and {Sgr{\`o}}, C. and {Siskind}, J. and {Smith}, D.~A. and {Smith}, P.~D. and {Spandre}, G. and {Spinelli}, P. and {Starck}, J.-L. and {Stawarz}, L. and {Strickman}, M.~S. and {Suson}, D.~J. and {Tajima}, H. and {Takahashi}, H. and {Takahashi}, T. and {Tanaka}, T. and {Thayer}, J.~B. and {Thayer}, J.~G. and {Thompson}, D.~J. and {Tibaldo}, L. and {Torres}, D.~F. and {Tosti}, G. and {Tramacere}, A. and {Uchiyama}, T.~L., Y. Usher and {Vasileiou}, V. and {Vilchez}, N. and {Vitale}, V. and {Waite}, A.~P. and {Wallace}, E. and {Wang}, P. and {Winer}, B.~L. and {Wood}, K.~S. and {Ylinen}, T. and {Ziegler}, M. and {Hardcastle}, M.~J. and {Kazanas}, D. and {Fermi LAT Collaboration}},
        title = "{Fermi Gamma-Ray Imaging of a Radio Galaxy}",
      journal = {Science},
         year = 2010,
        month = may,
       volume = {328},
       number = {5979},
        pages = {725},
          doi = {10.1126/science.1184656},
archivePrefix = {arXiv},
       eprint = {1006.3986},
 primaryClass = {astro-ph.HE},
       adsurl = {https://ui.adsabs.harvard.edu/abs/2010Sci...328..725A}
}

@ARTICLE{chilufya25mnras,
       author = {{Chilufya}, J. and {Hardcastle}, M.~J. and {Pierce}, J.~C.~S. and {Drake}, A.~B. and {Baldi}, R.~D. and {R{\"o}ttgering}, H.~J.~A. and {Smith}, D.~J.~B.},
        title = "{The nature of HERGs and LERGs in LoTSS DR2 - a morphological perspective}",
      journal = {\mnras},
         year = 2025,
        month = may,
       volume = {539},
       number = {1},
        pages = {463-486},
          doi = {10.1093/mnras/staf508},
archivePrefix = {arXiv},
       eprint = {2503.20586},
 primaryClass = {astro-ph.GA},
       adsurl = {https://ui.adsabs.harvard.edu/abs/2025MNRAS.539..463C}
}

@ARTICLE{cid10mnras,
       author = {{Cid Fernandes}, R. and {Stasi{\'n}ska}, G. and {Schlickmann}, M.~S. and {Mateus}, A. and {Vale Asari}, N. and {Schoenell}, W. and {Sodr{\'e}}, L.},
        title = "{Alternative diagnostic diagrams and the `forgotten' population of weak line galaxies in the SDSS}",
      journal = {\mnras},
         year = 2010,
        month = apr,
       volume = {403},
       number = {2},
        pages = {1036-1053},
          doi = {10.1111/j.1365-2966.2009.16185.x},
archivePrefix = {arXiv},
       eprint = {0912.1643},
 primaryClass = {astro-ph.CO},
       adsurl = {https://ui.adsabs.harvard.edu/abs/2010MNRAS.403.1036C}
}

@ARTICLE{kauffmann03apj,
       author = {{Kauffmann}, Guinevere and {Heckman}, Timothy M. and {Tremonti}, Christy and {Brinchmann}, Jarle and {Charlot}, St{\'e}phane and {White}, Simon D.~M. and {Ridgway}, Susan E. and {Brinkmann}, Jon and {Fukugita}, Masataka and {Hall}, Patrick B. and {Ivezi{\'c}}, {\v{Z}}eljko and {Richards}, Gordon T. and {Schneider}, Donald P.},
        title = "{The host galaxies of active galactic nuclei}",
      journal = {\mnras},
         year = 2003,
        month = dec,
       volume = {346},
       number = {4},
        pages = {1055-1077},
          doi = {10.1111/j.1365-2966.2003.07154.x},
archivePrefix = {arXiv},
       eprint = {astro-ph/0304239},
 primaryClass = {astro-ph},
       adsurl = {https://ui.adsabs.harvard.edu/abs/2003MNRAS.346.1055K}
}

@ARTICLE{kewley2001apj,
       author = {{Kewley}, L.~J. and {Dopita}, M.~A. and {Sutherland}, R.~S. and {Heisler}, C.~A. and {Trevena}, J.},
        title = "{Theoretical Modeling of Starburst Galaxies}",
      journal = {\apj},
         year = 2001,
        month = jul,
       volume = {556},
       number = {1},
        pages = {121-140},
          doi = {10.1086/321545},
archivePrefix = {arXiv},
       eprint = {astro-ph/0106324},
 primaryClass = {astro-ph},
       adsurl = {https://ui.adsabs.harvard.edu/abs/2001ApJ...556..121K}
}

@ARTICLE{paliya25apj,
       author = {{Paliya}, Vaidehi S. and {Saikia}, D.~J. and {Bruni}, G. and {Dom{\'\i}nguez}, Alberto and {Stalin}, C.~S.},
        title = "{Radio Morphology of Gamma-Ray Sources. II. Giant Radio Galaxies}",
      journal = {\apj},
         year = 2025,
        month = aug,
       volume = {989},
       number = {1},
          eid = {36},
        pages = {36},
          doi = {10.3847/1538-4357/adef0c},
archivePrefix = {arXiv},
       eprint = {2507.03105},
 primaryClass = {astro-ph.HE},
       adsurl = {https://ui.adsabs.harvard.edu/abs/2025ApJ...989...36P}
}

@ARTICLE{paliya24apj,
       author = {{Paliya}, Vaidehi S. and {Saikia}, D.~J. and {Dom{\'\i}nguez}, Alberto and {Stalin}, C.~S.},
        title = "{Radio Morphology of Gamma-Ray Sources: Double-lobed Radio Sources}",
      journal = {\apj},
         year = 2024,
        month = nov,
       volume = {976},
       number = {1},
          eid = {120},
        pages = {120},
          doi = {10.3847/1538-4357/ad85e2},
archivePrefix = {arXiv},
       eprint = {2410.10192},
 primaryClass = {astro-ph.HE},
       adsurl = {https://ui.adsabs.harvard.edu/abs/2024ApJ...976..120P}
}

@ARTICLE{makarov14AA,
       author = {{Makarov}, Dmitry and {Prugniel}, Philippe and {Terekhova}, Nataliya and {Courtois}, H{\'e}l{\`e}ne and {Vauglin}, Isabelle},
        title = "{HyperLEDA. III. The catalogue of extragalactic distances}",
      journal = {\aap},
         year = 2014,
        month = oct,
       volume = {570},
          eid = {A13},
        pages = {A13},
          doi = {10.1051/0004-6361/201423496},
archivePrefix = {arXiv},
       eprint = {1408.3476},
 primaryClass = {astro-ph.GA},
       adsurl = {https://ui.adsabs.harvard.edu/abs/2014A&A...570A..13M}
}

@ARTICLE{lewis06apj,
       author = {{Lewis}, Karen T. and {Eracleous}, Michael},
        title = "{Black Hole Masses of Active Galaxies with Double-peaked Balmer Emission Lines}",
      journal = {\apj},
         year = 2006,
        month = may,
       volume = {642},
       number = {2},
        pages = {711-719},
          doi = {10.1086/501419},
archivePrefix = {arXiv},
       eprint = {astro-ph/0601398},
 primaryClass = {astro-ph},
       adsurl = {https://ui.adsabs.harvard.edu/abs/2006ApJ...642..711L}
}

@ARTICLE{hlabathe2020mnras,
       author = {{Hlabathe}, Michael S. and {Starkey}, David A. and {Horne}, Keith and {Romero-Colmenero}, Encarni and {Crawford}, Steven M. and {Valenti}, Stefano and {Winkler}, Hartmut and {Barth}, Aaron J. and {Onken}, Christopher A. and {Sand}, David J. and {Treu}, Tommaso and {Diamond-Stanic}, Aleksandar M. and {Villforth}, Carolin},
        title = "{Robotic reverberation mapping of the broad-line radio galaxy 3C 120}",
      journal = {\mnras},
         year = 2020,
        month = sep,
       volume = {497},
       number = {3},
        pages = {2910-2929},
          doi = {10.1093/mnras/staa2171},
archivePrefix = {arXiv},
       eprint = {2007.11522},
 primaryClass = {astro-ph.GA},
       adsurl = {https://ui.adsabs.harvard.edu/abs/2020MNRAS.497.2910H}
}

@ARTICLE{shields72apj,
       author = {{Shields}, G.~A. and {Oke}, J.~B. and {Sargent}, W.~L.~W.},
        title = "{The Optical Spectrum of the Seyfert Galaxy 3c 120}",
      journal = {\apj},
         year = 1972,
        month = aug,
       volume = {176},
        pages = {75},
          doi = {10.1086/151610},
       adsurl = {https://ui.adsabs.harvard.edu/abs/1972ApJ...176...75S}
}

@ARTICLE{eracleous03apj,
       author = {{Eracleous}, Michael and {Halpern}, Jules P.},
        title = "{Completion of a Survey and Detailed Study of Double-peaked Emission Lines in Radio-loud Active Galactic Nuclei}",
      journal = {\apj},
         year = 2003,
        month = dec,
       volume = {599},
       number = {2},
        pages = {886-908},
          doi = {10.1086/379540},
archivePrefix = {arXiv},
       eprint = {astro-ph/0309149},
 primaryClass = {astro-ph},
       adsurl = {https://ui.adsabs.harvard.edu/abs/2003ApJ...599..886E}
}

@ARTICLE{kim03apj,
       author = {{Kim}, Dong-Woo and {Fabbiano}, G.},
        title = "{Chandra X-Ray Observations of NGC 1316 (Fornax A)}",
      journal = {\apj},
         year = 2003,
        month = apr,
       volume = {586},
       number = {2},
        pages = {826-849},
          doi = {10.1086/367930},
archivePrefix = {arXiv},
       eprint = {astro-ph/0206369},
 primaryClass = {astro-ph},
       adsurl = {https://ui.adsabs.harvard.edu/abs/2003ApJ...586..826K}
}

@ARTICLE{baum92,
       author = {{Baum}, S.~A. and {Heckman}, T.~M. and {van Breugel}, W.},
        title = "{Spectroscopy of Emission-Line Nebulae in Powerful Radio Galaxies: Interpretation}",
      journal = {\apj},
         year = 1992,
        month = apr,
       volume = {389},
        pages = {208},
          doi = {10.1086/171198},
       adsurl = {https://ui.adsabs.harvard.edu/abs/1992ApJ...389..208B}
}

@ARTICLE{tremaine2002ApJ,
       author = {{Tremaine}, Scott and {Gebhardt}, Karl and {Bender}, Ralf and {Bower}, Gary and {Dressler}, Alan and {Faber}, S.~M. and {Filippenko}, Alexei V. and {Green}, Richard and {Grillmair}, Carl and {Ho}, Luis C. and {Kormendy}, John and {Lauer}, Tod R. and {Magorrian}, John and {Pinkney}, Jason and {Richstone}, Douglas},
        title = "{The Slope of the Black Hole Mass versus Velocity Dispersion Correlation}",
      journal = {\apj},
         year = 2002,
        month = aug,
       volume = {574},
       number = {2},
        pages = {740-753},
          doi = {10.1086/341002},
archivePrefix = {arXiv},
       eprint = {astro-ph/0203468},
 primaryClass = {astro-ph},
       adsurl = {https://ui.adsabs.harvard.edu/abs/2002ApJ...574..740T}
}

@ARTICLE{abdo.2010.apj.720.misagn,
       author = {{Abdo}, A.~A. and {Ackermann}, M. and {Ajello}, M. and {Baldini}, L. and {Ballet}, J. and {Barbiellini}, G. and {Bastieri}, D. and {Bechtol}, K. and {Bellazzini}, R. and {Berenji}, B. and {Blandford}, R.~D. and {Bloom}, E.~D. and {Bonamente}, E. and {Borgland}, A.~W. and {Bouvier}, A. and {Brandt}, T.~J. and {Bregeon}, J. and {Brez}, A. and {Brigida}, M. and {Bruel}, P. and {Buehler}, R. and {Burnett}, T.~H. and {Buson}, S. and {Caliandro}, G.~A. and {Cameron}, R.~A. and {Cannon}, A. and {Caraveo}, P.~A. and {Carrigan}, S. and {Casandjian}, J.~M. and {Cavazzuti}, E. and {Cecchi}, C. and {{\c{C}}elik}, {\"O}. and {Celotti}, A. and {Charles}, E. and {Chekhtman}, A. and {Chen}, A.~W. and {Cheung}, C.~C. and {Chiang}, J. and {Ciprini}, S. and {Claus}, R. and {Cohen-Tanugi}, J. and {Colafrancesco}, S. and {Conrad}, J. and {Davis}, D.~S. and {Dermer}, C.~D. and {de Angelis}, A. and {de Palma}, F. and {Silva}, E. do Couto e. and {Drell}, P.~S. and {Dubois}, R. and {Favuzzi}, C. and {Fegan}, S.~J. and {Ferrara}, E.~C. and {Fortin}, P. and {Frailis}, M. and {Fukazawa}, Y. and {Fusco}, P. and {Gargano}, F. and {Gasparrini}, D. and {Gehrels}, N. and {Germani}, S. and {Giglietto}, N. and {Giommi}, P. and {Giordano}, F. and {Giroletti}, M. and {Glanzman}, T. and {Godfrey}, G. and {Grandi}, P. and {Grenier}, I.~A. and {Grove}, J.~E. and {Guillemot}, L. and {Guiriec}, S. and {Hadasch}, D. and {Hayashida}, M. and {Hays}, E. and {Horan}, D. and {Hughes}, R.~E. and {Jackson}, M.~S. and {J{\'o}hannesson}, G. and {Johnson}, A.~S. and {Johnson}, W.~N. and {Kamae}, T. and {Katagiri}, H. and {Kataoka}, J. and {Kn{\"o}dlseder}, J. and {Kuss}, M. and {Lande}, J. and {Latronico}, L. and {Lee}, S. -H. and {Lemoine-Goumard}, M. and {Llena Garde}, M. and {Longo}, F. and {Loparco}, F. and {Lott}, B. and {Lovellette}, M.~N. and {Lubrano}, P. and {Madejski}, G.~M. and {Makeev}, A. and {Malaguti}, G. and {Mazziotta}, M.~N. and {McConville}, W. and {McEnery}, J.~E. and {Michelson}, P.~F. and {Migliori}, G. and {Mitthumsiri}, W. and {Mizuno}, T. and {Monte}, C. and {Monzani}, M.~E. and {Morselli}, A. and {Moskalenko}, I.~V. and {Murgia}, S. and {Naumann-Godo}, M. and {Nestoras}, I. and {Nolan}, P.~L. and {Norris}, J.~P. and {Nuss}, E. and {Ohsugi}, T. and {Okumura}, A. and {Omodei}, N. and {Orlando}, E. and {Ormes}, J.~F. and {Paneque}, D. and {Panetta}, J.~H. and {Parent}, D. and {Pelassa}, V. and {Pepe}, M. and {Persic}, M. and {Pesce-Rollins}, M. and {Piron}, F. and {Porter}, T.~A. and {Rain{\`o}}, S. and {Rando}, R. and {Razzano}, M. and {Razzaque}, S. and {Reimer}, A. and {Reimer}, O. and {Reyes}, L.~C. and {Roth}, M. and {Sadrozinski}, H.~F. -W. and {Sanchez}, D. and {Sander}, A. and {Scargle}, J.~D. and {Sgr{\`o}}, C. and {Siskind}, E.~J. and {Smith}, P.~D. and {Spandre}, G. and {Spinelli}, P. and {Stawarz}, {\L}. and {Stecker}, F.~W. and {Strickman}, M.~S. and {Suson}, D.~J. and {Takahashi}, H. and {Tanaka}, T. and {Thayer}, J.~B. and {Thayer}, J.~G. and {Thompson}, D.~J. and {Tibaldo}, L. and {Torres}, D.~F. and {Torresi}, E. and {Tosti}, G. and {Tramacere}, A. and {Uchiyama}, Y. and {Usher}, T.~L. and {Vandenbroucke}, J. and {Vasileiou}, V. and {Vilchez}, N. and {Villata}, M. and {Vitale}, V. and {Waite}, A.~P. and {Wang}, P. and {Winer}, B.~L. and {Wood}, K.~S. and {Yang}, Z. and {Ylinen}, T. and {Ziegler}, M.},
        title = "{Fermi Large Area Telescope Observations of Misaligned Active Galactic Nuclei}",
      journal = {\apj},
         year = 2010,
        month = sep,
       volume = {720},
       number = {1},
        pages = {912-922},
          doi = {10.1088/0004-637X/720/1/912},
archivePrefix = {arXiv},
       eprint = {1007.1624},
 primaryClass = {astro-ph.HE},
       adsurl = {https://ui.adsabs.harvard.edu/abs/2010ApJ...720..912A}
}

@ARTICLE{bassi18,
       author = {{Bassi}, T. and {Migliori}, G. and {Grandi}, P. and {Vignali}, C. and {P{\'e}rez-Torres}, M.~A. and {Baldi}, R.~D. and {Torresi}, E. and {Siemiginowska}, A. and {Stanghellini}, C.},
        title = "{Faint {\ensuremath{\gamma}}-ray sources at low redshift: the radio galaxy IC 1531}",
      journal = {\mnras},
         year = 2018,
        month = dec,
       volume = {481},
       number = {4},
        pages = {5236-5246},
          doi = {10.1093/mnras/sty2622},
archivePrefix = {arXiv},
       eprint = {1810.02668},
 primaryClass = {astro-ph.HE},
       adsurl = {https://ui.adsabs.harvard.edu/abs/2018MNRAS.481.5236B}
}

@ARTICLE{abdo10sed,
       author = {{Abdo}, A.~A. and {Ackermann}, M. and {Agudo}, I. and {Ajello}, M. and {Aller}, H.~D. and {Aller}, M.~F. and {Angelakis}, E. and {Arkharov}, A.~A. and {Axelsson}, M. and {Bach}, U. and {Baldini}, L. and {Ballet}, J. and {Barbiellini}, G. and {Bastieri}, D. and {Baughman}, B.~M. and {Bechtol}, K. and {Bellazzini}, R. and {Benitez}, E. and {Berdyugin}, A. and {Berenji}, B. and {Blandford}, R.~D. and {Bloom}, E.~D. and {Boettcher}, M. and {Bonamente}, E. and {Borgland}, A.~W. and {Bregeon}, J. and {Brez}, A. and {Brigida}, M. and {Bruel}, P. and {Burnett}, T.~H. and {Burrows}, D. and {Buson}, S. and {Caliandro}, G.~A. and {Calzoletti}, L. and {Cameron}, R.~A. and {Capalbi}, M. and {Caraveo}, P.~A. and {Carosati}, D. and {Casandjian}, J.~M. and {Cavazzuti}, E. and {Cecchi}, C. and {{\c{C}}elik}, {\"O}. and {Charles}, E. and {Chaty}, S. and {Chekhtman}, A. and {Chen}, W.~P. and {Chiang}, J. and {Chincarini}, G. and {Ciprini}, S. and {Claus}, R. and {Cohen-Tanugi}, J. and {Colafrancesco}, S. and {Cominsky}, L.~R. and {Conrad}, J. and {Costamante}, L. and {Cutini}, S. and {D'ammando}, F. and {Deitrick}, R. and {D'Elia}, V. and {Dermer}, C.~D. and {de Angelis}, A. and {de Palma}, F. and {Digel}, S.~W. and {Donnarumma}, I. and {Silva}, E. do Couto e. and {Drell}, P.~S. and {Dubois}, R. and {Dultzin}, D. and {Dumora}, D. and {Falcone}, A. and {Farnier}, C. and {Favuzzi}, C. and {Fegan}, S.~J. and {Focke}, W.~B. and {Forn{\'e}}, E. and {Fortin}, P. and {Frailis}, M. and {Fuhrmann}, L. and {Fukazawa}, Y. and {Funk}, S. and {Fusco}, P. and {G{\'o}mez}, J.~L. and {Gargano}, F. and {Gasparrini}, D. and {Gehrels}, N. and {Germani}, S. and {Giebels}, B. and {Giglietto}, N. and {Giommi}, P. and {Giordano}, F. and {Giuliani}, A. and {Glanzman}, T. and {Godfrey}, G. and {Grenier}, I.~A. and {Gronwall}, C. and {Grove}, J.~E. and {Guillemot}, L. and {Guiriec}, S. and {Gurwell}, M.~A. and {Hadasch}, D. and {Hanabata}, Y. and {Harding}, A.~K. and {Hayashida}, M. and {Hays}, E. and {Healey}, S.~E. and {Heidt}, J. and {Hiriart}, D. and {Horan}, D. and {Hoversten}, E.~A. and {Hughes}, R.~E. and {Itoh}, R. and {Jackson}, M.~S. and {J{\'o}hannesson}, G. and {Johnson}, A.~S. and {Johnson}, W.~N. and {Jorstad}, S.~G. and {Kadler}, M. and {Kamae}, T. and {Katagiri}, H. and {Kataoka}, J. and {Kawai}, N. and {Kennea}, J. and {Kerr}, M. and {Kimeridze}, G. and {Kn{\"o}dlseder}, J. and {Kocian}, M.~L. and {Kopatskaya}, E.~N. and {Koptelova}, E. and {Konstantinova}, T.~S. and {Kovalev}, Y.~Y. and {Kovalev}, Yu. A. and {Kurtanidze}, O.~M. and {Kuss}, M. and {Lande}, J. and {Larionov}, V.~M. and {Latronico}, L. and {Leto}, P. and {Lindfors}, E. and {Longo}, F. and {Loparco}, F. and {Lott}, B. and {Lovellette}, M.~N. and {Lubrano}, P. and {Madejski}, G.~M. and {Makeev}, A. and {Marchegiani}, P. and {Marscher}, A.~P. and {Marshall}, F. and {Max-Moerbeck}, W. and {Mazziotta}, M.~N. and {McConville}, W. and {McEnery}, J.~E. and {Meurer}, C. and {Michelson}, P.~F. and {Mitthumsiri}, W. and {Mizuno}, T. and {Moiseev}, A.~A. and {Monte}, C. and {Monzani}, M.~E. and {Morselli}, A. and {Moskalenko}, I.~V. and {Murgia}, S. and {Nestoras}, I. and {Nilsson}, K. and {Nizhelsky}, N.~A. and {Nolan}, P.~L. and {Norris}, J.~P. and {Nuss}, E. and {Ohsugi}, T. and {Ojha}, R. and {Omodei}, N. and {Orlando}, E. and {Ormes}, J.~F. and {Osborne}, J. and {Ozaki}, M. and {Pacciani}, L. and {Padovani}, P. and {Pagani}, C. and {Page}, K. and {Paneque}, D. and {Panetta}, J.~H. and {Parent}, D. and {Pasanen}, M. and {Pavlidou}, V. and {Pelassa}, V. and {Pepe}, M. and {Perri}, M. and {Pesce-Rollins}, M. and {Piranomonte}, S. and {Piron}, F. and {Pittori}, C. and {Porter}, T.~A. and {Puccetti}, S. and {Rahoui}, F. and {Rain{\`o}}, S. and {Raiteri}, C. and {Rando}, R. and {Razzano}, M. and {Reimer}, A. and {Reimer}, O.},
        title = "{The Spectral Energy Distribution of Fermi Bright Blazars}",
      journal = {\apj},
         year = 2010,
        month = jun,
       volume = {716},
       number = {1},
        pages = {30-70},
          doi = {10.1088/0004-637X/716/1/30},
archivePrefix = {arXiv},
       eprint = {0912.2040},
 primaryClass = {astro-ph.CO},
       adsurl = {https://ui.adsabs.harvard.edu/abs/2010ApJ...716...30A}
}

@ARTICLE{fossati98mnras,
       author = {{Fossati}, G. and {Maraschi}, L. and {Celotti}, A. and {Comastri}, A. and {Ghisellini}, G.},
        title = "{A unifying view of the spectral energy distributions of blazars}",
      journal = {\mnras},
         year = 1998,
        month = sep,
       volume = {299},
       number = {2},
        pages = {433-448},
          doi = {10.1046/j.1365-8711.1998.01828.x},
archivePrefix = {arXiv},
       eprint = {astro-ph/9804103},
 primaryClass = {astro-ph},
       adsurl = {https://ui.adsabs.harvard.edu/abs/1998MNRAS.299..433F}
}

@ARTICLE{mingo14mnras,
       author = {{Mingo}, B. and {Hardcastle}, M.~J. and {Croston}, J.~H. and {Dicken}, D. and {Evans}, D.~A. and {Morganti}, R. and {Tadhunter}, C.},
        title = "{An X-ray survey of the 2 Jy sample - I. Is there an accretion mode dichotomy in radio-loud AGN?}",
      journal = {\mnras},
         year = 2014,
        month = may,
       volume = {440},
       number = {1},
        pages = {269-297},
          doi = {10.1093/mnras/stu263},
archivePrefix = {arXiv},
       eprint = {1402.1770},
 primaryClass = {astro-ph.GA},
       adsurl = {https://ui.adsabs.harvard.edu/abs/2014MNRAS.440..269M}
}

@ARTICLE{morganti1999AA,
       author = {{Morganti}, R. and {Oosterloo}, T. and {Tadhunter}, C.~N. and {Aiudi}, R. and {Jones}, P. and {Villar-Martin}, M.},
        title = "{The radio structures of southern 2-Jy radio sources: New ATCA and VLA radio images}",
      journal = {\aaps},
         year = 1999,
        month = dec,
       volume = {140},
        pages = {355-372},
          doi = {10.1051/aas:1999427},
archivePrefix = {arXiv},
       eprint = {astro-ph/9910150},
 primaryClass = {astro-ph},
       adsurl = {https://ui.adsabs.harvard.edu/abs/1999A&AS..140..355M}
}

@ARTICLE{fan16,
       author = {{Fan}, J.~H. and {Yang}, J.~H. and {Liu}, Y. and {Luo}, G.~Y. and {Lin}, C. and {Yuan}, Y.~H. and {Xiao}, H.~B. and {Zhou}, A.~Y. and {Hua}, T.~X. and {Pei}, Z.~Y.},
        title = "{The Spectral Energy Distributions of Fermi Blazars}",
      journal = {\apjs},
         year = 2016,
        month = oct,
       volume = {226},
       number = {2},
          eid = {20},
        pages = {20},
          doi = {10.3847/0067-0049/226/2/20},
archivePrefix = {arXiv},
       eprint = {1608.03958},
 primaryClass = {astro-ph.HE},
       adsurl = {https://ui.adsabs.harvard.edu/abs/2016ApJS..226...20F}
}

@ARTICLE{xie12pasj,
       author = {{Xie}, Zhao Hua and {Zhang}, Li},
        title = "{The Jet-Disk Symbiosis in Blazar and Unified Model of FSRQs and BL Lac Objects}",
      journal = {\pasj},
         year = 2012,
        month = apr,
       volume = {64},
       number = {2},
          eid = {33},
        pages = {33},
          doi = {10.1093/pasj/64.2.33},
       adsurl = {https://ui.adsabs.harvard.edu/abs/2012PASJ...64...33X}
}

@INPROCEEDINGS{grandi12ijmps,
       author = {{Grandi}, Paola},
        title = "{Gamma Rays from Radio Galaxies: FERMI/LAT Observations}",
    booktitle = {International Journal of Modern Physics Conference Series},
         year = 2012,
       series = {International Journal of Modern Physics Conference Series},
       volume = {8},
        month = jan,
        pages = {25-30},
          doi = {10.1142/S2010194512004370},
archivePrefix = {arXiv},
       eprint = {1112.2505},
 primaryClass = {astro-ph.HE},
       adsurl = {https://ui.adsabs.harvard.edu/abs/2012IJMPS...8...25G}
}

@ARTICLE{maruo14apj,
       author = {{Di Mauro}, M. and {Calore}, F. and {Donato}, F. and {Ajello}, M. and {Latronico}, L.},
        title = "{Diffuse {\ensuremath{\gamma}}-Ray Emission from Misaligned Active Galactic Nuclei}",
      journal = {\apj},
         year = 2014,
        month = jan,
       volume = {780},
       number = {2},
          eid = {161},
        pages = {161},
          doi = {10.1088/0004-637X/780/2/161},
archivePrefix = {arXiv},
       eprint = {1304.0908},
 primaryClass = {astro-ph.HE},
       adsurl = {https://ui.adsabs.harvard.edu/abs/2014ApJ...780..161D}
}

@ARTICLE{fu23raa,
       author = {{Fu}, Wen-Jing and {Zhang}, Hai-Ming and {Zhang}, Jin and {Liang}, Yun-Feng and {Yao}, Su and {Liang}, En-Wei},
        title = "{Is TOL 1326-379 a Prototype of {\ensuremath{\gamma}}-Ray Emitting FR0 Radio Galaxy?}",
      journal = {Research in Astronomy and Astrophysics},
         year = 2022,
        month = mar,
       volume = {22},
       number = {3},
          eid = {035005},
        pages = {035005},
          doi = {10.1088/1674-4527/ac4410},
       adsurl = {https://ui.adsabs.harvard.edu/abs/2022RAA....22c5005F}
}

@ARTICLE{Tadhunter08,
   author = {{Tadhunter}, C.},
    title = "{An introduction to active galactic nuclei: Classification and unification}",
  journal = {New Astronomy Reviews},
     year = 2008,
    month = aug,
   volume = 52,
    pages = {227-239},
      doi = {10.1016/j.newar.2008.06.004},
   adsurl = {http://adsabs.harvard.edu/abs/2008NewAR..52..227T}
}

@ARTICLE{foschini12raa,
       author = {{Foschini}, Luigi},
        title = "{On the emission lines in active galactic nuclei with relativistic jets}",
      journal = {Research in Astronomy and Astrophysics},
         year = 2012,
        month = apr,
       volume = {12},
       number = {4},
        pages = {359-368},
          doi = {10.1088/1674-4527/12/4/001},
archivePrefix = {arXiv},
       eprint = {1103.2008},
 primaryClass = {astro-ph.GA},
       adsurl = {https://ui.adsabs.harvard.edu/abs/2012RAA....12..359F}
}

@ARTICLE{fan2013PASJ,
       author = {{Fan}, Junhui and {Yang}, Jiang He and {Zhang}, Jing-Yi and {Hua}, Tong Xu and {Liu}, Yi and {Qin}, Yi-Ping and {Huang}, Yong},
        title = "{Beaming Effect in Fermi Blazars}",
      journal = {\pasj},
         year = 2013,
        month = apr,
       volume = {65},
          eid = {25},
        pages = {25},
          doi = {10.1093/pasj/65.2.25},
archivePrefix = {arXiv},
       eprint = {1210.4096},
 primaryClass = {astro-ph.HE},
       adsurl = {https://ui.adsabs.harvard.edu/abs/2013PASJ...65...25F}
}

@ARTICLE{macconi2020mnras,
       author = {{Macconi}, D. and {Torresi}, E. and {Grandi}, P. and {Boccardi}, B. and {Vignali}, C.},
        title = "{Radio morphology-accretion mode link in Fanaroff-Riley type II low-excitation radio galaxies}",
      journal = {\mnras},
         year = 2020,
        month = apr,
       volume = {493},
       number = {3},
        pages = {4355-4366},
          doi = {10.1093/mnras/staa560},
archivePrefix = {arXiv},
       eprint = {2002.09360},
 primaryClass = {astro-ph.HE},
       adsurl = {https://ui.adsabs.harvard.edu/abs/2020MNRAS.493.4355M}
}

@ARTICLE{boccardi2021aa,
       author = {{Boccardi}, B. and {Perucho}, M. and {Casadio}, C. and {Grandi}, P. and {Macconi}, D. and {Torresi}, E. and {Pellegrini}, S. and {Krichbaum}, T.~P. and {Kadler}, M. and {Giovannini}, G. and {Karamanavis}, V. and {Ricci}, L. and {Madika}, E. and {Bach}, U. and {Ros}, E. and {Giroletti}, M. and {Zensus}, J.~A.},
        title = "{Jet collimation in NGC 315 and other nearby AGN}",
      journal = {\aap},
         year = 2021,
        month = mar,
       volume = {647},
          eid = {A67},
        pages = {A67},
          doi = {10.1051/0004-6361/202039612},
archivePrefix = {arXiv},
       eprint = {2012.14831},
 primaryClass = {astro-ph.HE},
       adsurl = {https://ui.adsabs.harvard.edu/abs/2021A&A...647A..67B}
}

@ARTICLE{ineson2015MNRAS,
       author = {{Ineson}, J. and {Croston}, J.~H. and {Hardcastle}, M.~J. and {Kraft}, R.~P. and {Evans}, D.~A. and {Jarvis}, M.},
        title = "{The link between accretion mode and environment in radio-loud active galaxies}",
      journal = {\mnras},
         year = 2015,
        month = nov,
       volume = {453},
       number = {3},
        pages = {2682-2706},
          doi = {10.1093/mnras/stv1807},
archivePrefix = {arXiv},
       eprint = {1508.01033},
 primaryClass = {astro-ph.GA},
       adsurl = {https://ui.adsabs.harvard.edu/abs/2015MNRAS.453.2682I}
}

@ARTICLE{sadler2014MNRAS,
       author = {{Sadler}, Elaine M. and {Ekers}, Ronald D. and {Mahony}, Elizabeth K. and {Mauch}, Tom and {Murphy}, Tara},
        title = "{The local radio-galaxy population at 20 GHz}",
      journal = {\mnras},
         year = 2014,
        month = feb,
       volume = {438},
       number = {1},
        pages = {796-824},
          doi = {10.1093/mnras/stt2239},
archivePrefix = {arXiv},
       eprint = {1304.0268},
 primaryClass = {astro-ph.CO},
       adsurl = {https://ui.adsabs.harvard.edu/abs/2014MNRAS.438..796S}
}

@INPROCEEDINGS{ghisellini2011aipc,
       author = {{Ghisellini}, Gabriele},
        title = "{Extragalactic relativistic jets}",
    booktitle = {25th Texas Symposium on Relativistic AstroPhysics (Texas 2010)},
         year = 2011,
       editor = {{Aharonian}, Felix A. and {Hofmann}, Werner and {Rieger}, Frank M.},
       series = {American Institute of Physics Conference Series},
       volume = {1381},
        month = sep,
    publisher = {AIP},
        pages = {180-198},
          doi = {10.1063/1.3635832},
archivePrefix = {arXiv},
       eprint = {1104.0006},
 primaryClass = {astro-ph.CO},
       adsurl = {https://ui.adsabs.harvard.edu/abs/2011AIPC.1381..180G}
}

@ARTICLE{hardcastle07,
       author = {{Hardcastle}, M.~J. and {Evans}, D.~A. and {Croston}, J.~H.},
        title = "{Hot and cold gas accretion and feedback in radio-loud active galaxies}",
      journal = {\mnras},
         year = 2007,
        month = apr,
       volume = {376},
       number = {4},
        pages = {1849-1856},
          doi = {10.1111/j.1365-2966.2007.11572.x},
archivePrefix = {arXiv},
       eprint = {astro-ph/0701857},
 primaryClass = {astro-ph},
       adsurl = {https://ui.adsabs.harvard.edu/abs/2007MNRAS.376.1849H}
}

@ARTICLE{baldi10b,
       author = {{Baldi}, Ranieri D. and {Chiaberge}, Marco and {Capetti}, Alessandro and {Sparks}, William and {Macchetto}, F. Duccio and {O'Dea}, Christopher P. and {Axon}, David J. and {Baum}, Stefi A. and {Quillen}, Alice C.},
        title = "{The 1.6 {\ensuremath{\mu}}m Near-infrared Nuclei of 3C Radio Galaxies: Jets, Thermal Emission, or Scattered Light?}",
      journal = {\apj},
         year = 2010,
        month = dec,
       volume = {725},
       number = {2},
        pages = {2426-2443},
          doi = {10.1088/0004-637X/725/2/2426},
archivePrefix = {arXiv},
       eprint = {1010.5277},
 primaryClass = {astro-ph.CO},
       adsurl = {https://ui.adsabs.harvard.edu/abs/2010ApJ...725.2426B}
}

@ARTICLE{mingo22,
       author = {{Mingo}, B. and {Croston}, J.~H. and {Best}, P.~N. and {Duncan}, K.~J. and {Hardcastle}, M.~J. and {Kondapally}, R. and {Prandoni}, I. and {Sabater}, J. and {Shimwell}, T.~W. and {Williams}, W.~L. and {Baldi}, R.~D. and {Bonato}, M. and {Bondi}, M. and {Dabhade}, P. and {G{\"u}rkan}, G. and {Ineson}, J. and {Magliocchetti}, M. and {Miley}, G. and {Pierce}, J.~C.~S. and {R{\"o}ttgering}, H.~J.~A.},
        title = "{Accretion mode versus radio morphology in the LOFAR Deep Fields}",
      journal = {\mnras},
         year = 2022,
        month = apr,
       volume = {511},
       number = {3},
        pages = {3250-3271},
          doi = {10.1093/mnras/stac140},
archivePrefix = {arXiv},
       eprint = {2201.04433},
 primaryClass = {astro-ph.GA},
       adsurl = {https://ui.adsabs.harvard.edu/abs/2022MNRAS.511.3250M}
}

@ARTICLE{baldi23,
       author = {{Baldi}, Ranieri D.},
        title = "{The nature of compact radio sources: the case of FR 0 radio galaxies}",
      journal = {\aapr},
         year = 2023,
        month = dec,
       volume = {31},
       number = {1},
          eid = {3},
        pages = {3},
          doi = {10.1007/s00159-023-00148-3},
archivePrefix = {arXiv},
       eprint = {2307.08379},
 primaryClass = {astro-ph.GA},
       adsurl = {https://ui.adsabs.harvard.edu/abs/2023A&ARv..31....3B}
}

@ARTICLE{khatiya2024apj,
       author = {{Khatiya}, Nikita S. and {Boughelilba}, Margot and {Karwin}, Christopher M. and {McDaniel}, Alex and {Zhao}, Xiurui and {Ajello}, Marco and {Reimer}, Anita and {Hartmann}, Dieter H.},
        title = "{Characterizing the {\ensuremath{\gamma}}-Ray Emission from FR0 Radio Galaxies}",
      journal = {\apj},
         year = 2024,
        month = aug,
       volume = {971},
       number = {1},
          eid = {84},
        pages = {84},
          doi = {10.3847/1538-4357/ad534c},
archivePrefix = {arXiv},
       eprint = {2310.19888},
 primaryClass = {astro-ph.HE},
       adsurl = {https://ui.adsabs.harvard.edu/abs/2024ApJ...971...84K}
}

@ARTICLE{angioni17,
       author = {{Angioni}, R. and {Grandi}, P. and {Torresi}, E. and {Vignali}, C. and {Kn{\"o}dlseder}, J.},
        title = "{Radio galaxies with the Cherenkov Telescope Array}",
      journal = {Astroparticle Physics},
         year = 2017,
        month = jun,
       volume = {92},
        pages = {42-48},
          doi = {10.1016/j.astropartphys.2017.02.010},
archivePrefix = {arXiv},
       eprint = {1702.05926},
 primaryClass = {astro-ph.HE},
       adsurl = {https://ui.adsabs.harvard.edu/abs/2017APh....92...42A}
}

@ARTICLE{abdo10b,
       author = {{Abdo}, A.~A. and {Ackermann}, M. and {Ajello}, M. and {Baldini}, L. and {Ballet}, J. and {Barbiellini}, G. and {Bastieri}, D. and {Bechtol}, K. and {Bellazzini}, R. and {Berenji}, B. and {Blandford}, R.~D. and {Bloom}, E.~D. and {Bonamente}, E. and {Borgland}, A.~W. and {Bouvier}, A. and {Brandt}, T.~J. and {Bregeon}, J. and {Brez}, A. and {Brigida}, M. and {Bruel}, P. and {Buehler}, R. and {Burnett}, T.~H. and {Buson}, S. and {Caliandro}, G.~A. and {Cameron}, R.~A. and {Cannon}, A. and {Caraveo}, P.~A. and {Carrigan}, S. and {Casandjian}, J.~M. and {Cavazzuti}, E. and {Cecchi}, C. and {{\c{C}}elik}, {\"O}. and {Celotti}, A. and {Charles}, E. and {Chekhtman}, A. and {Chen}, A.~W. and {Cheung}, C.~C. and {Chiang}, J. and {Ciprini}, S. and {Claus}, R. and {Cohen-Tanugi}, J. and {Colafrancesco}, S. and {Conrad}, J. and {Davis}, D.~S. and {Dermer}, C.~D. and {de Angelis}, A. and {de Palma}, F. and {Silva}, E. do Couto e. and {Drell}, P.~S. and {Dubois}, R. and {Favuzzi}, C. and {Fegan}, S.~J. and {Ferrara}, E.~C. and {Fortin}, P. and {Frailis}, M. and {Fukazawa}, Y. and {Fusco}, P. and {Gargano}, F. and {Gasparrini}, D. and {Gehrels}, N. and {Germani}, S. and {Giglietto}, N. and {Giommi}, P. and {Giordano}, F. and {Giroletti}, M. and {Glanzman}, T. and {Godfrey}, G. and {Grandi}, P. and {Grenier}, I.~A. and {Grove}, J.~E. and {Guillemot}, L. and {Guiriec}, S. and {Hadasch}, D. and {Hayashida}, M. and {Hays}, E. and {Horan}, D. and {Hughes}, R.~E. and {Jackson}, M.~S. and {J{\'o}hannesson}, G. and {Johnson}, A.~S. and {Johnson}, W.~N. and {Kamae}, T. and {Katagiri}, H. and {Kataoka}, J. and {Kn{\"o}dlseder}, J. and {Kuss}, M. and {Lande}, J. and {Latronico}, L. and {Lee}, S. -H. and {Lemoine-Goumard}, M. and {Llena Garde}, M. and {Longo}, F. and {Loparco}, F. and {Lott}, B. and {Lovellette}, M.~N. and {Lubrano}, P. and {Madejski}, G.~M. and {Makeev}, A. and {Malaguti}, G. and {Mazziotta}, M.~N. and {McConville}, W. and {McEnery}, J.~E. and {Michelson}, P.~F. and {Migliori}, G. and {Mitthumsiri}, W. and {Mizuno}, T. and {Monte}, C. and {Monzani}, M.~E. and {Morselli}, A. and {Moskalenko}, I.~V. and {Murgia}, S. and {Naumann-Godo}, M. and {Nestoras}, I. and {Nolan}, P.~L. and {Norris}, J.~P. and {Nuss}, E. and {Ohsugi}, T. and {Okumura}, A. and {Omodei}, N. and {Orlando}, E. and {Ormes}, J.~F. and {Paneque}, D. and {Panetta}, J.~H. and {Parent}, D. and {Pelassa}, V. and {Pepe}, M. and {Persic}, M. and {Pesce-Rollins}, M. and {Piron}, F. and {Porter}, T.~A. and {Rain{\`o}}, S. and {Rando}, R. and {Razzano}, M. and {Razzaque}, S. and {Reimer}, A. and {Reimer}, O. and {Reyes}, L.~C. and {Roth}, M. and {Sadrozinski}, H.~F. -W. and {Sanchez}, D. and {Sander}, A. and {Scargle}, J.~D. and {Sgr{\`o}}, C. and {Siskind}, E.~J. and {Smith}, P.~D. and {Spandre}, G. and {Spinelli}, P. and {Stawarz}, {\L}. and {Stecker}, F.~W. and {Strickman}, M.~S. and {Suson}, D.~J. and {Takahashi}, H. and {Tanaka}, T. and {Thayer}, J.~B. and {Thayer}, J.~G. and {Thompson}, D.~J. and {Tibaldo}, L. and {Torres}, D.~F. and {Torresi}, E. and {Tosti}, G. and {Tramacere}, A. and {Uchiyama}, Y. and {Usher}, T.~L. and {Vandenbroucke}, J. and {Vasileiou}, V. and {Vilchez}, N. and {Villata}, M. and {Vitale}, V. and {Waite}, A.~P. and {Wang}, P. and {Winer}, B.~L. and {Wood}, K.~S. and {Yang}, Z. and {Ylinen}, T. and {Ziegler}, M.},
        title = "{Fermi Large Area Telescope Observations of Misaligned Active Galactic Nuclei}",
      journal = {\apj},
         year = 2010,
        month = sep,
       volume = {720},
       number = {1},
        pages = {912-922},
          doi = {10.1088/0004-637X/720/1/912},
archivePrefix = {arXiv},
       eprint = {1007.1624},
 primaryClass = {astro-ph.HE},
       adsurl = {https://ui.adsabs.harvard.edu/abs/2010ApJ...720..912A}
}

@ARTICLE{grandi25,
       author = {{Grandi}, P. and {Giovannini}, G. and {Torresi}, E. and {Boccardi}, B.},
        title = "{Peering into the heart of 3CR radio galaxies: A very long baseline interferometry perspective on optical-radio classifications at parsec scales}",
      journal = {\aap},
         year = 2025,
        month = jul,
       volume = {699},
          eid = {A286},
        pages = {A286},
          doi = {10.1051/0004-6361/202453018},
archivePrefix = {arXiv},
       eprint = {2506.07589},
 primaryClass = {astro-ph.HE},
       adsurl = {https://ui.adsabs.harvard.edu/abs/2025A&A...699A.286G}
}

@ARTICLE{baldi10,
       author = {{Baldi}, R.~D. and {Capetti}, A.},
        title = "{Spectro-photometric properties of the bulk of the radio-loud AGN population}",
      journal = {\aap},
         year = 2010,
        month = sep,
       volume = {519},
          eid = {A48},
        pages = {A48},
          doi = {10.1051/0004-6361/201014446},
archivePrefix = {arXiv},
       eprint = {1005.3223},
 primaryClass = {astro-ph.CO},
       adsurl = {https://ui.adsabs.harvard.edu/abs/2010A&A...519A..48B}
}

@ARTICLE{ghisellini01aa,
       author = {{Ghisellini}, G. and {Celotti}, A.},
        title = "{The dividing line between FR I and FR II radio-galaxies}",
      journal = {\aap},
         year = 2001,
        month = nov,
       volume = {379},
        pages = {L1-L4},
          doi = {10.1051/0004-6361:20011338},
archivePrefix = {arXiv},
       eprint = {astro-ph/0106570},
 primaryClass = {astro-ph},
       adsurl = {https://ui.adsabs.harvard.edu/abs/2001A&A...379L...1G}
}

@ARTICLE{zhao24apj,
       author = {{Zhao}, X.~Z. and {Yang}, H.~Y. and {Zheng}, Y.~G. and {Kang}, S.~J.},
        title = "{The Energy Budget in the Jet of High-frequency Peaked BL Lacertae Objects}",
      journal = {\apj},
         year = 2024,
        month = jun,
       volume = {967},
       number = {2},
          eid = {104},
        pages = {104},
          doi = {10.3847/1538-4357/ad3ba9},
archivePrefix = {arXiv},
       eprint = {2406.01046},
 primaryClass = {astro-ph.HE},
       adsurl = {https://ui.adsabs.harvard.edu/abs/2024ApJ...967..104Z}
}

@ARTICLE{capetti02aa,
       author = {{Capetti}, A. and {Celotti}, A. and {Chiaberge}, M. and {de Ruiter}, H.~R. and {Fanti}, R. and {Morganti}, R. and {Parma}, P.},
        title = "{The HST survey of the B2 sample of radio-galaxies: Optical nuclei and the FR I/BL Lac unified scheme}",
      journal = {\aap},
         year = 2002,
        month = jan,
       volume = {383},
        pages = {104-111},
          doi = {10.1051/0004-6361:20011714},
archivePrefix = {arXiv},
       eprint = {astro-ph/0112151},
 primaryClass = {astro-ph},
       adsurl = {https://ui.adsabs.harvard.edu/abs/2002A&A...383..104C}
}

@ARTICLE{ghisellini11mnras,
       author = {{Ghisellini}, G. and {Tavecchio}, F. and {Foschini}, L. and {Ghirlanda}, G.},
        title = "{The transition between BL Lac objects and flat spectrum radio quasars}",
      journal = {\mnras},
         year = 2011,
        month = jul,
       volume = {414},
       number = {3},
        pages = {2674-2689},
          doi = {10.1111/j.1365-2966.2011.18578.x},
archivePrefix = {arXiv},
       eprint = {1012.0308},
 primaryClass = {astro-ph.CO},
       adsurl = {https://ui.adsabs.harvard.edu/abs/2011MNRAS.414.2674G}
}

@ARTICLE{sbarrato14mnras,
       author = {{Sbarrato}, T. and {Padovani}, P. and {Ghisellini}, G.},
        title = "{The jet-disc connection in AGN}",
      journal = {\mnras},
         year = 2014,
        month = nov,
       volume = {445},
       number = {1},
        pages = {81-92},
          doi = {10.1093/mnras/stu1759},
archivePrefix = {arXiv},
       eprint = {1405.4865},
 primaryClass = {astro-ph.HE},
       adsurl = {https://ui.adsabs.harvard.edu/abs/2014MNRAS.445...81S}
}

@ARTICLE{buttiglione10aa,
       author = {{Buttiglione}, S. and {Capetti}, A. and {Celotti}, A. and {Axon}, D.~J. and {Chiaberge}, M. and {Macchetto}, F.~D. and {Sparks}, W.~B.},
        title = "{An optical spectroscopic survey of the 3CR sample of radio galaxies with z < 0.3 . II. Spectroscopic classes and accretion modes in radio-loud AGN}",
      journal = {\aap},
         year = 2010,
        month = jan,
       volume = {509},
          eid = {A6},
        pages = {A6},
          doi = {10.1051/0004-6361/200913290},
archivePrefix = {arXiv},
       eprint = {0911.0536},
 primaryClass = {astro-ph.CO},
       adsurl = {https://ui.adsabs.harvard.edu/abs/2010A&A...509A...6B}
}

@ARTICLE{keenan21mn,
       author = {{Keenan}, Mary and {Meyer}, Eileen T. and {Georganopoulos}, Markos and {Reddy}, Karthik and {French}, Omar J.},
        title = "{The relativistic jet dichotomy and the end of the blazar sequence}",
      journal = {\mnras},
         year = 2021,
        month = aug,
       volume = {505},
       number = {4},
        pages = {4726-4745},
          doi = {10.1093/mnras/stab1182},
archivePrefix = {arXiv},
       eprint = {2007.12661},
 primaryClass = {astro-ph.GA},
       adsurl = {https://ui.adsabs.harvard.edu/abs/2021MNRAS.505.4726K}
}

@ARTICLE{chen15aj,
       author = {{Chen}, Yong-Yun and {Zhang}, Xiong and {Xiong}, Dingrong and {Yu}, Xiaoling},
        title = "{Black Hole Mass, Jet Power, and Accretion in AGNs}",
      journal = {\aj},
         year = 2015,
        month = jul,
       volume = {150},
       number = {1},
          eid = {8},
        pages = {8},
          doi = {10.1088/0004-6256/150/1/8},
archivePrefix = {arXiv},
       eprint = {1504.05413},
 primaryClass = {astro-ph.HE},
       adsurl = {https://ui.adsabs.harvard.edu/abs/2015AJ....150....8C}
}

@INPROCEEDINGS{lai94,
       author = {{Laing}, R.~A. and {Jenkins}, C.~R. and {Wall}, J.~V. and {Unger}, S.~W.},
        title = "{Spectrophotometry of a Complete Sample of 3CR Radio Sources: Implications for Unified Models}",
    booktitle = {The Physics of Active Galaxies},
         year = 1994,
       editor = {{Bicknell}, Geoffrey V. and {Dopita}, Michael A. and {Quinn}, Peter J.},
       series = {Astronomical Society of the Pacific Conference Series},
       volume = {54},
        month = jan,
        pages = {201},
       adsurl = {https://ui.adsabs.harvard.edu/abs/1994ASPC...54..201L}
}

@ARTICLE{capetti05,
       author = {{Capetti}, A. and {Verdoes Kleijn}, G. and {Chiaberge}, M.},
        title = "{The HST view of the nuclear emission line region in low luminosity radio-galaxies}",
      journal = {\aap},
         year = 2005,
        month = sep,
       volume = {439},
       number = {3},
        pages = {935-946},
          doi = {10.1051/0004-6361:20041609},
archivePrefix = {arXiv},
       eprint = {astro-ph/0504568},
 primaryClass = {astro-ph},
       adsurl = {https://ui.adsabs.harvard.edu/abs/2005A&A...439..935C}
}

@ARTICLE{arnaudova25,
       author = {{Arnaudova}, M.~I. and {Smith}, D.~J.~B. and {Hardcastle}, M.~J. and {Best}, P.~N. and {Das}, S. and {Shenoy}, S. and {Duncan}, K.~J. and {Holden}, L.~R. and {Kondapally}, R. and {Morabito}, L.~K. and {R{\"o}ttgering}, H.~J.~A.},
        title = "{The LOFAR Two-metre Sky Survey Deep Fields: new probabilistic spectroscopic classifications and the accretion rates of radio galaxies}",
      journal = {\mnras},
         year = 2025,
        month = sep,
       volume = {542},
       number = {3},
        pages = {2245-2268},
          doi = {10.1093/mnras/staf1347},
archivePrefix = {arXiv},
       eprint = {2508.18347},
 primaryClass = {astro-ph.GA},
       adsurl = {https://ui.adsabs.harvard.edu/abs/2025MNRAS.542.2245A}
}

@ARTICLE{boula26,
       author = {{Boula}, S. and {Mastichiadis}, A. and {Kazanas}, D.},
        title = "{A one-parameter two-zone leptonic model for the blazar sequence}",
      journal = {\aap},
         year = 2026,
        month = jan,
       volume = {705},
          eid = {L5},
        pages = {L5},
          doi = {10.1051/0004-6361/202557600},
archivePrefix = {arXiv},
       eprint = {2512.14479},
 primaryClass = {astro-ph.HE},
       adsurl = {https://ui.adsabs.harvard.edu/abs/2026A&A...705L...5B}
}

@ARTICLE{grandi21,
       author = {{Grandi}, Paola and {Torresi}, Eleonora and {Macconi}, Duccio and {Boccardi}, Bia and {Capetti}, Alessandro},
        title = "{Jet-Accretion System in the Nearby mJy Radio Galaxies}",
      journal = {\apj},
         year = 2021,
        month = apr,
       volume = {911},
       number = {1},
          eid = {17},
        pages = {17},
          doi = {10.3847/1538-4357/abe776},
archivePrefix = {arXiv},
       eprint = {2102.08922},
 primaryClass = {astro-ph.HE},
       adsurl = {https://ui.adsabs.harvard.edu/abs/2021ApJ...911...17G}
}

@ARTICLE{baldi19b,
       author = {{Baldi}, Ranieri D. and {Capetti}, Alessandro and {Giovannini}, Gabriele},
        title = "{High-resolution VLA observations of FR0 radio galaxies: the properties and nature of compact radio sources}",
      journal = {\mnras},
         year = 2019,
        month = jan,
       volume = {482},
       number = {2},
        pages = {2294-2304},
          doi = {10.1093/mnras/sty2703},
archivePrefix = {arXiv},
       eprint = {1810.01894},
 primaryClass = {astro-ph.GA},
       adsurl = {https://ui.adsabs.harvard.edu/abs/2019MNRAS.482.2294B}
}

@ARTICLE{baldi19,
       author = {{Baldi}, Ranieri Diego and {Torresi}, Eleonora and {Migliori}, Giulia and {Balmaverde}, Barbara},
        title = "{The High Energy View of FR0 Radio Galaxies}",
      journal = {Galaxies},
         year = 2019,
        month = sep,
       volume = {7},
       number = {3},
          eid = {76},
        pages = {76},
          doi = {10.3390/galaxies7030076},
archivePrefix = {arXiv},
       eprint = {1909.04113},
 primaryClass = {astro-ph.HE},
       adsurl = {https://ui.adsabs.harvard.edu/abs/2019Galax...7...76B}
}

@ARTICLE{pannikkote23apj,
       author = {{Pannikkote}, Meghana and {Paliya}, Vaidehi S. and {Saikia}, D.~J.},
        title = "{Hunting Gamma-Ray-emitting FR0 Radio Galaxies in Wide-field Sky Surveys}",
      journal = {\apj},
         year = 2023,
        month = nov,
       volume = {957},
       number = {2},
          eid = {73},
        pages = {73},
          doi = {10.3847/1538-4357/ad00b5},
archivePrefix = {arXiv},
       eprint = {2310.03321},
 primaryClass = {astro-ph.HE},
       adsurl = {https://ui.adsabs.harvard.edu/abs/2023ApJ...957...73P}
}

@INPROCEEDINGS{torresi22MnSAI,
       author = {{Torresi}, E. and {Balmaverde}, B. and {Liuzzo}, E. and {Giovannini}, G. and {Paladino}, R. and {Baldi}, R.~D. and {Boccardi}, B. and {Capetti}, A. and {Ciprini}, S. and {Dadina}, M. and {D'Ammando}, F. and {Gasparrini}, D. and {Giroletti}, M. and {Grandi}, P. and {Lico}, R. and {Macconi}, D. and {Migliori}, G. and {Prandoni}, I. and {Raiteri}, C.~M. and {Ruffa}, I. and {Vignali}, C.},
        title = "{Exploring the radio morphology-accretion mode link in radio galaxies at high energies}",
    booktitle = {Memorie della Societa Astronomica Italiana},
         year = 2022,
       volume = {93},
        month = nov,
        pages = {81},
          doi = {10.36116/MEMSAIT_93N2_3.2022.10},
       adsurl = {https://ui.adsabs.harvard.edu/abs/2022MmSAI..93b..81T}
}

@ARTICLE{meyer2011apj,
       author = {{Meyer}, Eileen T. and {Fossati}, Giovanni and {Georganopoulos}, Markos and {Lister}, Matthew L.},
        title = "{From the Blazar Sequence to the Blazar Envelope: Revisiting the Relativistic Jet Dichotomy in Radio-loud Active Galactic Nuclei}",
      journal = {\apj},
         year = 2011,
        month = oct,
       volume = {740},
       number = {2},
          eid = {98},
        pages = {98},
          doi = {10.1088/0004-637X/740/2/98},
archivePrefix = {arXiv},
       eprint = {1107.5105},
 primaryClass = {astro-ph.CO},
       adsurl = {https://ui.adsabs.harvard.edu/abs/2011ApJ...740...98M}
}

@ARTICLE{huang99apj,
       author = {{Huang}, L.~H. and {Jiang}, D.~R. and {Cao}, Xinwu},
        title = "{A correlation analysis of flux ratios and the Doppler factor for EGRET AGN sources}",
      journal = {\aap},
         year = 1999,
        month = jan,
       volume = {341},
        pages = {74-80},
          doi = {10.48550/arXiv.astro-ph/9809251},
archivePrefix = {arXiv},
       eprint = {astro-ph/9809251},
 primaryClass = {astro-ph},
       adsurl = {https://ui.adsabs.harvard.edu/abs/1999A&A...341...74H}
}

@ARTICLE{massaro20apjl,
       author = {{Massaro}, F. and {Capetti}, A. and {Paggi}, A. and {Baldi}, R.~D. and {Tramacere}, A. and {Pillitteri}, I. and {Campana}, R.},
        title = "{Dragon's Lair: On the Large-scale Environment of BL Lac Objects}",
      journal = {\apjl},
         year = 2020,
        month = sep,
       volume = {900},
       number = {2},
          eid = {L34},
        pages = {L34},
          doi = {10.3847/2041-8213/abac56},
archivePrefix = {arXiv},
       eprint = {2009.03318},
 primaryClass = {astro-ph.GA},
       adsurl = {https://ui.adsabs.harvard.edu/abs/2020ApJ...900L..34M}
}

@ARTICLE{paliya21apjl,
       author = {{Paliya}, Vaidehi S.},
        title = "{A New Gamma-Ray-emitting Population of FR0 Radio Galaxies}",
      journal = {\apjl},
         year = 2021,
        month = sep,
       volume = {918},
       number = {2},
          eid = {L39},
        pages = {L39},
          doi = {10.3847/2041-8213/ac2143},
archivePrefix = {arXiv},
       eprint = {2108.11701},
 primaryClass = {astro-ph.HE},
       adsurl = {https://ui.adsabs.harvard.edu/abs/2021ApJ...918L..39P}
}

@ARTICLE{grandi16mnras,
       author = {{Grandi}, Paola and {Capetti}, Alessandro and {Baldi}, Ranieri D.},
        title = "{Discovery of a Fanaroff-Riley type 0 radio galaxy emitting at {\ensuremath{\gamma}}-ray energies}",
      journal = {\mnras},
         year = 2016,
        month = mar,
       volume = {457},
       number = {1},
        pages = {2-8},
          doi = {10.1093/mnras/stv2846},
archivePrefix = {arXiv},
       eprint = {1512.01242},
 primaryClass = {astro-ph.GA},
       adsurl = {https://ui.adsabs.harvard.edu/abs/2016MNRAS.457....2G}
}

@ARTICLE{homan.2021.apj,
       author = {{Homan}, D.~C. and {Cohen}, M.~H. and {Hovatta}, T. and {Kellermann}, K.~I. and {Kovalev}, Y.~Y. and {Lister}, M.~L. and {Popkov}, A.~V. and {Pushkarev}, A.~B. and {Ros}, E. and {Savolainen}, T.},
        title = "{MOJAVE. XIX. Brightness Temperatures and Intrinsic Properties of Blazar Jets}",
      journal = {\apj},
         year = 2021,
        month = dec,
       volume = {923},
       number = {1},
          eid = {67},
        pages = {67},
          doi = {10.3847/1538-4357/ac27af},
archivePrefix = {arXiv},
       eprint = {2109.04977},
 primaryClass = {astro-ph.HE},
       adsurl = {https://ui.adsabs.harvard.edu/abs/2021ApJ...923...67H}
}

@ARTICLE{capetti.02.nar,
       author = {{Capetti}, A. and {Trussoni}, E. and {Celotti}, A. and {Feretti}, L. and {Chiaberge}, M.},
        title = "{Spectral energy distributions of five FR I radio galaxies}",
      journal = {\nar},
         year = 2002,
        month = may,
       volume = {46},
       number = {2-7},
        pages = {335-337},
          doi = {10.1016/S1387-6473(01)00203-2},
archivePrefix = {arXiv},
       eprint = {astro-ph/0007434},
 primaryClass = {astro-ph},
       adsurl = {https://ui.adsabs.harvard.edu/abs/2002NewAR..46..335C}
}

@ARTICLE{ulgiati25,
       author = {{Ulgiati}, Alberto and {Padovani}, Paolo and {Giommi}, Paolo and {Paiano}, Simona and {Pinto}, Ciro},
        title = "{Characterization of a sample of {\ensuremath{\gamma}}-ray active galactic nuclei}",
      journal = {\mnras},
         year = 2025,
        month = oct,
       volume = {543},
       number = {1},
        pages = {326-350},
          doi = {10.1093/mnras/staf1459},
archivePrefix = {arXiv},
       eprint = {2509.02092},
 primaryClass = {astro-ph.HE},
       adsurl = {https://ui.adsabs.harvard.edu/abs/2025MNRAS.543..326U}
}

@ARTICLE{raimundo09,
       author = {{Raimundo}, S.~I. and {Fabian}, A.~C.},
        title = "{Eddington ratio and accretion efficiency in active galactic nuclei evolution}",
      journal = {\mnras},
         year = 2009,
        month = jul,
       volume = {396},
       number = {3},
        pages = {1217-1221},
          doi = {10.1111/j.1365-2966.2009.14796.x},
archivePrefix = {arXiv},
       eprint = {0903.3432},
 primaryClass = {astro-ph.GA},
       adsurl = {https://ui.adsabs.harvard.edu/abs/2009MNRAS.396.1217R}
}

@ARTICLE{lian25,
       author = {{Lian}, Ji-Shun and {Wang}, Ze-Rui and {Zhang}, Jin},
        title = "{Hadronic Processes in Advection-dominated Accretion Flow as the Origin of TeV Excesses in BL Lac Objects}",
      journal = {\apj},
         year = 2025,
        month = dec,
       volume = {995},
       number = {1},
          eid = {38},
        pages = {38},
          doi = {10.3847/1538-4357/ae1cc2},
archivePrefix = {arXiv},
       eprint = {2511.04202},
 primaryClass = {astro-ph.HE},
       adsurl = {https://ui.adsabs.harvard.edu/abs/2025ApJ...995...38L}
}

@ARTICLE{sbarrato12mnras,
       author = {{Sbarrato}, T. and {Ghisellini}, G. and {Maraschi}, L. and {Colpi}, M.},
        title = "{The relation between broad lines and {\ensuremath{\gamma}}-ray luminosities in Fermi blazars}",
      journal = {\mnras},
         year = 2012,
        month = apr,
       volume = {421},
       number = {2},
        pages = {1764-1778},
          doi = {10.1111/j.1365-2966.2012.20442.x},
archivePrefix = {arXiv},
       eprint = {1108.0927},
 primaryClass = {astro-ph.HE},
       adsurl = {https://ui.adsabs.harvard.edu/abs/2012MNRAS.421.1764S}
}

@ARTICLE{capetti.17.aa.fr2,
       author = {{Capetti}, A. and {Massaro}, F. and {Baldi}, R.~D.},
        title = "{FRIICAT: A FIRST catalog of FR II radio galaxies}",
      journal = {\aap},
         year = 2017,
        month = may,
       volume = {601},
          eid = {A81},
        pages = {A81},
          doi = {10.1051/0004-6361/201630247},
archivePrefix = {arXiv},
       eprint = {1703.03427},
 primaryClass = {astro-ph.HE},
       adsurl = {https://ui.adsabs.harvard.edu/abs/2017A&A...601A..81C}
}

@ARTICLE{capetti.17.aa.fr1,
       author = {{Capetti}, A. and {Massaro}, F. and {Baldi}, R.~D.},
        title = "{FRICAT: A FIRST catalog of FR I radio galaxies}",
      journal = {\aap},
         year = 2017,
        month = feb,
       volume = {598},
          eid = {A49},
        pages = {A49},
          doi = {10.1051/0004-6361/201629287},
archivePrefix = {arXiv},
       eprint = {1610.09376},
 primaryClass = {astro-ph.HE},
       adsurl = {https://ui.adsabs.harvard.edu/abs/2017A&A...598A..49C}
}

@ARTICLE{baldi.18.aa.fr0,
       author = {{Baldi}, R.~D. and {Capetti}, A. and {Massaro}, F.},
        title = "{FR0CAT: a FIRST catalog of FR 0 radio galaxies}",
      journal = {\aap},
         year = 2018,
        month = jan,
       volume = {609},
          eid = {A1},
        pages = {A1},
          doi = {10.1051/0004-6361/201731333},
archivePrefix = {arXiv},
       eprint = {1709.00015},
 primaryClass = {astro-ph.GA},
       adsurl = {https://ui.adsabs.harvard.edu/abs/2018A&A...609A...1B}
}

@ARTICLE{shakura73,
   author = {{Shakura}, N.~I. and {Sunyaev}, R.~A.},
    title = "{Black holes in binary systems. Observational appearance.}",
  journal = {\aap},
     year = 1973,
   volume = 24,
    pages = {337-355},
   adsurl = {http://adsabs.harvard.edu/abs/1973A%26A....24..337S}
}

@ARTICLE{abdollahi.apjs.2022.260,
       author = {{Abdollahi}, S. and {Acero}, F. and {Baldini}, L. and {Ballet}, J. and {Bastieri}, D. and {Bellazzini}, R. and {Berenji}, B. and {Berretta}, A. and {Bissaldi}, E. and {Blandford}, R.~D. and {Bloom}, E. and {Bonino}, R. and {Brill}, A. and {Britto}, R.~J. and {Bruel}, P. and {Burnett}, T.~H. and {Buson}, S. and {Cameron}, R.~A. and {Caputo}, R. and {Caraveo}, P.~A. and {Castro}, D. and {Chaty}, S. and {Cheung}, C.~C. and {Chiaro}, G. and {Cibrario}, N. and {Ciprini}, S. and {Coronado-Bl{\'a}zquez}, J. and {Crnogorcevic}, M. and {Cutini}, S. and {D'Ammando}, F. and {De Gaetano}, S. and {Digel}, S.~W. and {Di Lalla}, N. and {Dirirsa}, F. and {Di Venere}, L. and {Dom{\'\i}nguez}, A. and {Fallah Ramazani}, V. and {Fegan}, S.~J. and {Ferrara}, E.~C. and {Fiori}, A. and {Fleischhack}, H. and {Franckowiak}, A. and {Fukazawa}, Y. and {Funk}, S. and {Fusco}, P. and {Galanti}, G. and {Gammaldi}, V. and {Gargano}, F. and {Garrappa}, S. and {Gasparrini}, D. and {Giacchino}, F. and {Giglietto}, N. and {Giordano}, F. and {Giroletti}, M. and {Glanzman}, T. and {Green}, D. and {Grenier}, I.~A. and {Grondin}, M. -H. and {Guillemot}, L. and {Guiriec}, S. and {Gustafsson}, M. and {Harding}, A.~K. and {Hays}, E. and {Hewitt}, J.~W. and {Horan}, D. and {Hou}, X. and {J{\'o}hannesson}, G. and {Karwin}, C. and {Kayanoki}, T. and {Kerr}, M. and {Kuss}, M. and {Landriu}, D. and {Larsson}, S. and {Latronico}, L. and {Lemoine-Goumard}, M. and {Li}, J. and {Liodakis}, I. and {Longo}, F. and {Loparco}, F. and {Lott}, B. and {Lubrano}, P. and {Maldera}, S. and {Malyshev}, D. and {Manfreda}, A. and {Mart{\'\i}-Devesa}, G. and {Mazziotta}, M.~N. and {Mereu}, I. and {Meyer}, M. and {Michelson}, P.~F. and {Mirabal}, N. and {Mitthumsiri}, W. and {Mizuno}, T. and {Moiseev}, A.~A. and {Monzani}, M.~E. and {Morselli}, A. and {Moskalenko}, I.~V. and {Negro}, M. and {Nuss}, E. and {Omodei}, N. and {Orienti}, M. and {Orlando}, E. and {Paneque}, D. and {Pei}, Z. and {Perkins}, J.~S. and {Persic}, M. and {Pesce-Rollins}, M. and {Petrosian}, V. and {Pillera}, R. and {Poon}, H. and {Porter}, T.~A. and {Principe}, G. and {Rain{\`o}}, S. and {Rando}, R. and {Rani}, B. and {Razzano}, M. and {Razzaque}, S. and {Reimer}, A. and {Reimer}, O. and {Reposeur}, T. and {S{\'a}nchez-Conde}, M. and {Saz Parkinson}, P.~M. and {Scotton}, L. and {Serini}, D. and {Sgr{\`o}}, C. and {Siskind}, E.~J. and {Smith}, D.~A. and {Spandre}, G. and {Spinelli}, P. and {Sueoka}, K. and {Suson}, D.~J. and {Tajima}, H. and {Tak}, D. and {Thayer}, J.~B. and {Thompson}, D.~J. and {Torres}, D.~F. and {Troja}, E. and {Valverde}, J. and {Wood}, K. and {Zaharijas}, G.},
        title = "{Incremental Fermi Large Area Telescope Fourth Source Catalog}",
      journal = {\apjs},
         year = 2022,
        month = jun,
       volume = {260},
       number = {2},
          eid = {53},
        pages = {53},
          doi = {10.3847/1538-4365/ac6751},
archivePrefix = {arXiv},
       eprint = {2201.11184},
 primaryClass = {astro-ph.HE},
       adsurl = {https://ui.adsabs.harvard.edu/abs/2022ApJS..260...53A}
}

@ARTICLE{ajello.2022.apjs.263,
       author = {{Ajello}, M. and {Baldini}, L. and {Ballet}, J. and {Bastieri}, D. and {Becerra Gonzalez}, J. and {Bellazzini}, R. and {Berretta}, A. and {Bissaldi}, E. and {Bonino}, R. and {Brill}, A. and {Bruel}, P. and {Buson}, S. and {Caputo}, R. and {Caraveo}, P.~A. and {Cheung}, C.~C. and {Chiaro}, G. and {Cibrario}, N. and {Ciprini}, S. and {Crnogorcevic}, M. and {Cutini}, S. and {D'Ammando}, F. and {De Gaetano}, S. and {Di Lalla}, N. and {Di Venere}, L. and {Dom{\'\i}nguez}, A. and {Ramazani}, V. Fallah and {Ferrara}, E.~C. and {Fiori}, A. and {Fukazawa}, Y. and {Funk}, S. and {Fusco}, P. and {Gammaldi}, V. and {Gargano}, F. and {Garrappa}, S. and {Gasparrini}, D. and {Giglietto}, N. and {Giordano}, F. and {Giroletti}, M. and {Green}, D. and {Grenier}, I.~A. and {Guiriec}, S. and {Horan}, D. and {Hou}, X. and {Kayanoki}, T. and {Kuss}, M. and {Larsson}, S. and {Latronico}, L. and {Lewis}, T. and {Li}, J. and {Liodakis}, I. and {Longo}, F. and {Loparco}, F. and {Lott}, B. and {Lovellette}, M.~N. and {Lubrano}, P. and {Madejski}, G.~M. and {Maldera}, S. and {Manfreda}, A. and {Mart{\'\i}-Devesa}, G. and {Mazziotta}, M.~N. and {Mereu}, I. and {Michelson}, P.~F. and {Mirabal}, N. and {Mitthumsiri}, W. and {Mizuno}, T. and {Monzani}, M.~E. and {Morselli}, A. and {Moskalenko}, I.~V. and {Negro}, M. and {Ojha}, R. and {Orienti}, M. and {Orlando}, E. and {Ormes}, J.~F. and {Pei}, Z. and {Pe{\~n}a-Herazo}, H. and {Persic}, M. and {Pesce-Rollins}, M. and {Petrosian}, V. and {Pillera}, R. and {Poon}, H. and {Porter}, T.~A. and {Principe}, G. and {Rain{\`o}}, S. and {Rando}, R. and {Rani}, B. and {Razzano}, M. and {Razzaque}, S. and {Reimer}, A. and {Reimer}, O. and {Scotton}, L. and {Serini}, D. and {Sgr{\`o}}, C. and {Siskind}, E.~J. and {Spandre}, G. and {Spinelli}, P. and {Suson}, D.~J. and {Tajima}, H. and {Torres}, D.~F. and {Valverde}, J. and {Yassin}, H. and {Zaharijas}, G.},
        title = "{The Fourth Catalog of Active Galactic Nuclei Detected by the Fermi Large Area Telescope: Data Release 3}",
      journal = {\apjs},
         year = 2022,
        month = dec,
       volume = {263},
       number = {2},
          eid = {24},
        pages = {24},
          doi = {10.3847/1538-4365/ac9523},
archivePrefix = {arXiv},
       eprint = {2209.12070},
 primaryClass = {astro-ph.HE},
       adsurl = {https://ui.adsabs.harvard.edu/abs/2022ApJS..263...24A}
}

%



\begin{appendix}


 \FloatBarrier 

 \end{appendix}
\end{document}